\RequirePackage{lineno}
\documentclass[aps,prb,11pt,tightenlines,superscriptaddress]{revtex4}

\usepackage{graphicx}
\usepackage{braket}
\usepackage{epstopdf}
\usepackage{color}
\usepackage{subfigure}
\usepackage{amsmath} 
\usepackage{amssymb} 
\usepackage{ulem}

\def\emph{\textit}

\begin{document}

\title{Surface spin magnetism controls the polarized exciton emission from CdSe nanoplatelets}

\author{Elena V. Shornikova}
\affiliation{Experimentelle Physik 2, Technische Universit{\"a}t Dortmund, 44221 Dortmund, Germany}
\email{elena.shornikova@tu-dortmund.de}
\author{Aleksandr A. Golovatenko}
\affiliation{Ioffe  Institute, Russian Academy of Sciences, 194021 St. Petersburg, Russia}
\author{Dmitri R. Yakovlev}
\affiliation{Experimentelle Physik 2, Technische Universit{\"a}t Dortmund, 44221 Dortmund, Germany}
\affiliation{Ioffe  Institute, Russian Academy of Sciences, 194021 St. Petersburg, Russia}
\email{dmitri.yakovlev@tu-dortmund.de}
\author{Anna V. Rodina}
\affiliation{Ioffe  Institute, Russian Academy of Sciences, 194021 St. Petersburg, Russia}
\author{Louis Biadala}
\affiliation{Institut d'Electronique, de Micro{\'e}lectronique et de Nanotechnologie, CNRS, 59652 Villeneuve-d'Ascq, France}
\author{Gang Qiang}
\affiliation{Experimentelle Physik 2, Technische Universit{\"a}t Dortmund, 44221 Dortmund, Germany}
\author{Alexis Kuntzmann}
\affiliation{Laboratoire de Physique et d'Etude des Mat\'{e}riaux, ESPCI, CNRS, 75231 Paris, France}
\author{Michel Nasilowski}
\affiliation{Laboratoire de Physique et d'Etude des Mat\'{e}riaux, ESPCI, CNRS, 75231 Paris, France}
\author{Benoit Dubertret}
\affiliation{Laboratoire de Physique et d'Etude des Mat\'{e}riaux, ESPCI, CNRS, 75231 Paris, France}
\author{Anatolii Polovitsyn}
\affiliation{Department of Chemistry, Ghent University, 9000 Ghent, Belgium}
\affiliation{Istituto Italiano di Tecnologia, 16163 Genova, Italy}
\author{Iwan Moreels}
\affiliation{Department of Chemistry, Ghent University, 9000 Ghent, Belgium}
\affiliation{Istituto Italiano di Tecnologia, 16163 Genova, Italy}
\author{Manfred Bayer}
\affiliation{Experimentelle Physik 2, Technische Universit{\"a}t Dortmund, 44221 Dortmund, Germany}
\affiliation{Ioffe  Institute, Russian Academy of Sciences, 194021 St. Petersburg, Russia}

\date{\today}

\begin{abstract}
The surface of nominally diamagnetic colloidal CdSe nanoplatelets can demonstrate paramagnetism owing to the uncompensated spins of dangling bonds (DBSs). We reveal that by optical spectroscopy in high magnetic fields up to 15 Tesla using the exciton spin as probe of the surface magnetism. The strongly nonlinear magnetic field dependence of the circular polarization of the exciton emission is determined by the DBS and exciton spin polarization as well as by the spin-dependent recombination of dark excitons. The sign of the exciton-DBS exchange interaction can be adjusted by the nanoplatelet growth conditions.     
\end{abstract}


\maketitle

The surface of colloidal nanocrystals (NCs) greatly controls their optical and electronic properties making the surface chemistry critically important in nanocrystal research and applications.\cite{Kovalenko2015,Pietryga2016,Nasilowski2016} Undercoordinated surface atoms with excess electrons, which in colloidal synthesis are often metal cations, either rearrange themselves by surface reconstruction or adsorb surfactant ligands.\cite{Owen2015,Drijvers2018} The ligands are used to control the colloidal synthesis, increase the NC solubility, screen the NCs from environment, and stabilize their surface by saturating dangling bonds of the surface atoms.\cite{Boles2016,Lorenzon2015}
They influence surface trap states and thereby control photoluminescence quantum yield.\cite{Owen2017,Singh2018,Meerbach2019,Ning2011}  Not every dangling bond can be passivated due to steric hindrance, as the ligand diameter typically exceeds the lattice constant of NC material and due to poor interaction of a facet with the ligands. An electron transfer from the $d$-shell of a surface atom to the ligand can provide surface magnetism.\cite{Crespo2004,Yamamoto2004,Meulenberg2009} Also the spins of unpassivated dangling bonds can contribute to it. The spins of surface atoms act similar to spins of magnetic impurities in diluted magnetic semiconductors,\cite{Furdyna1988,Beaulac2008,Delikanli2015} and may influence crucially the optical, electronic and magnetic properties of colloidal NCs.\cite{Rodina2015,Biadala2017} Nominally diamagnetic NCs  may demonstrate paramagnetic behavior and giant magneto-optical effects. 

Here, we study the surface spins in colloidal quasi-two-dimensional nanoplatelets (NPLs) based on CdSe semiconductor. These emerging nanostructures have an atomically controlled thickness of a few monolayers, providing remarkable optical properties with narrow emission lines of neutral and charged excitons.\cite{Ithurria2008,Shornikova2018ns} The interaction of confined excitons with the dangling-bond spins (DBSs) provides a nanoscopic tool for monitoring surface magnetism. In particular, the exciton spin polarization in magnetic field and the radiative recombination of dark excitons are strongly modified due to the exchange interaction with DBSs. We use high magnetic fields up to 15~T to measure the degree of circular polarization (DCP) of exciton photoluminescence and the exciton spin and recombination dynamics at cryogenic temperatures. In contract to a pure diamagnetic behavior, we find a strongly nonmonotonic magnetic field dependence of DCP, which sign changes for NPLs synthesized in air or argon atmosphere. This allows us to identify two mechanisms resulting from the exciton interaction with the surface spins. The first one is an additional Zeeman splitting of the exciton states, which is similar to the giant Zeeman splitting effect in diluted magnetic semiconductors. The second one is provided by the spin-dependent recombination of the dark excitons.\cite{Biadala2017,Rodina2018JEM}     
Beyond considering these two mechanisms in an external magnetic field, the developed model approach has to account for different NPL orientations relative to the field in the NC ensemble, to obtain an accurate modeling of the experimental data.

\textbf{Experimental results}

\begin{figure}[h!]
	\includegraphics{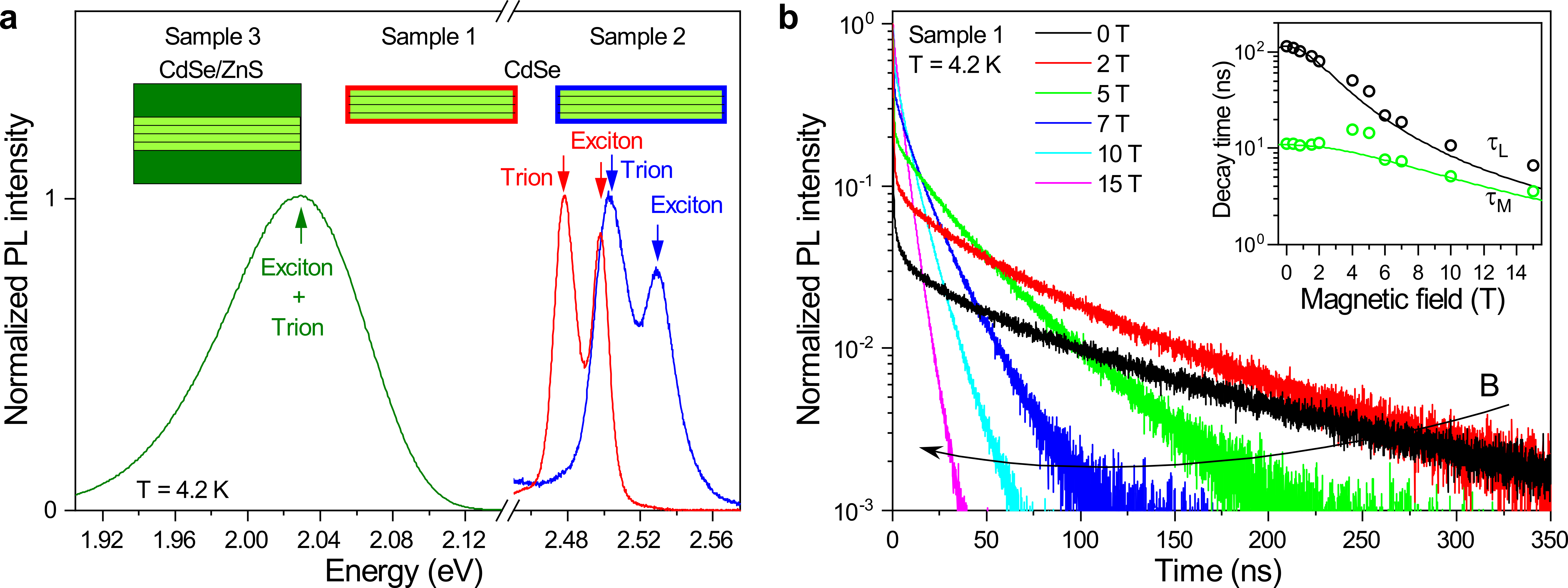}
	\caption{\textbf{PL spectra and recombination dynamics.} \textbf{a,} PL spectra of bare core CdSe NPLs (4 ML thickness), synthesized in argon (red) and in air (blue), and of CdSe/ZnS core/shell NPLs. The exciton and trion emission maxima are marked by arrows. Sample parameters are given in Table~\ref{tab:table1}. \textbf{b,} PL decays at the maximum of exciton line of Sample~1 measured at different magnetic fields. Inset: decay times, $\tau_{\rm M}$ and $\tau_{\rm L}$, as functions of magnetic field. Symbols are experimental data, lines are calculations according equation~\eqref{EQ:gammaFBM} using $\Delta E_{\rm AF}=4.6$~meV, $\Gamma_{\rm A}=10~\rm{ns}^{-1}$, $g_{\rm e}=1.7$ and fit parameter $\theta=\pi/2$. }
	\label{fig:Fig1ab}
\end{figure}

Low-temperature photoluminescence (PL) spectra of the studied NPLs are shown in Fig.~\ref{fig:Fig1ab}a. The bare core CdSe NPLs with 4 monolayers (MLs) thickness show two well-resolved emission lines corresponding to recombination of neutral and negatively charged excitons (trions).\cite{Shornikova2018ns}  The CdSe/ZnS NPLs has a broader peak red-shifted from the CdSe NPLs due to  electron spread into the shells. It is valuable in CdSe/CdS NPLs,\cite{Shornikova2018nl} but also relevant for CdSe/ZnS NPLs, despite the higher barriers.\cite{Cruguel2017} 

Figure~\ref{fig:Fig1ab}b shows the exciton recombination dynamics in Sample~1, for other samples see Supplementary Section~\ref{subsec:SI_TR(B)_Exper}. The short decay component, which dynamics is resolution limited to 0.3~ns, corresponds to the bright exciton lifetime contributed by its recombination and relaxation to the dark exciton. The long component $\tau_{\rm L}=114$~ns is the dark exciton emission.\cite{Biadala2014nl,Shornikova2018ns}  Accurate fitting  shows that in Samples~1 and 2 the exciton decay has three components (Supplementary Sections~\ref{subsec:SI_FStr_TR(T)_S1_S2},~\ref{subsec:SI_TR(B)_Theor}). The  middle decay time $\tau_{\rm M}=11$~ns  corresponds to NPLs with a finite nonradiative recombination rate $\Gamma_{\rm nr}$, which is nearly independent of temperature and magnetic field, so that $ 1/\tau_{\rm M}=1/\tau_{\rm L}+\Gamma_{\rm nr}$. The PL decay in Sample~3 is multi-exponential, evidencing that the broad PL band is contributed by exciton and trion emission (Supplementary Sections~\ref{subsec:SI_FStr_TR(T)_S3}).
The magnetic field dependences of $\tau_{\rm L}$ and  $\tau_{\rm M}$ in Sample 1  are shown in the inset of Fig.~\ref{fig:Fig1ab}b. Both times shorten with increasing field, which is explained by a shortening of the radiative recombination time of dark exciton due to admixture of bright exciton states\cite{Efros1996,Rodina2016} (Methods and Supplementary Section~\ref{subsec:SI_TR(B)_Theor}).

\begin{table*}
	\small
	\caption{\ Parameters of CdSe (Samples 1 and 2) and  CdSe/ZnS (Sample 3) nanoplatelets. $^a$ data taken from ref. \onlinecite{Shornikova2018ns}. }
	\begin{tabular*}{1\textwidth}{lccc}
		Sample & Sample 1 & Sample 2 & Sample 3\\ \hline
		Synthesis atmosphere & argon & air & argon\\
		CdSe thickness, ML & 4  & 4 & 4\\
		ZnS shell thickness, ML & -- & -- & 5 \\
		Lateral dimensions, nm$^2$ & $8 \times 16$$^a$ & $4.9 \times 24$ & $18 \times 18$ \\
		Emission photon energy ($T=300$~K), eV & 2.420 & 2.439 & 1.983\\
		Emission photon energy ($T=4.2$~K), eV & 2.497 & 2.529 & 2.025\\
		FWHM of exciton line ($T=4.2$~K), meV & 16.5 & 21.9 &92.5\\
		Shift between emission lines ($T=4.2$~K), meV & 20$^a$ & 26 & broad band\\
		Light-heavy hole splitting, meV & 157$^a$ & 146 & 220\\
		$\tau_{\rm M}$ ($T=4.2$~K, $B=0$~T), ns &11$\pm$1&11$\pm$1& -- \\
		$\tau_{\rm L}$ ($T=4.2$~K, $B=0$~T), ns &114$\pm$5&114$\pm$5& -- \\
		$\Gamma^0_{\rm F}=1/\tau_{\rm L}$ ($T=4.2$~K, $B=0$~T), ns$^{-1}$ &0.009&0.009& -- \\
		$\Gamma_{\rm nr}=1/\tau_{\rm M}-1/\tau_{\rm L}$, ns$^{-1}$ &0.083&0.083& -- \\
		$\Gamma_{\rm A}$, ns$^{-1}$ &10$^a$ &8& 0.4 \\
		$\Delta E_{\rm AF}$ (two-exponential fit), meV & $5.0 \pm 0.5$$^a$ &$5.0\pm 0.5$& $3.9\pm0.5$\\
		$\Delta E_{\rm AF}$ (three-exponential fit), meV & $4.6 \pm 0.5$ &$4.6\pm0.5$& --\\ \hline
	\end{tabular*}
	\label{tab:table1}
\end{table*}

\begin{figure*}[h!]
	\includegraphics{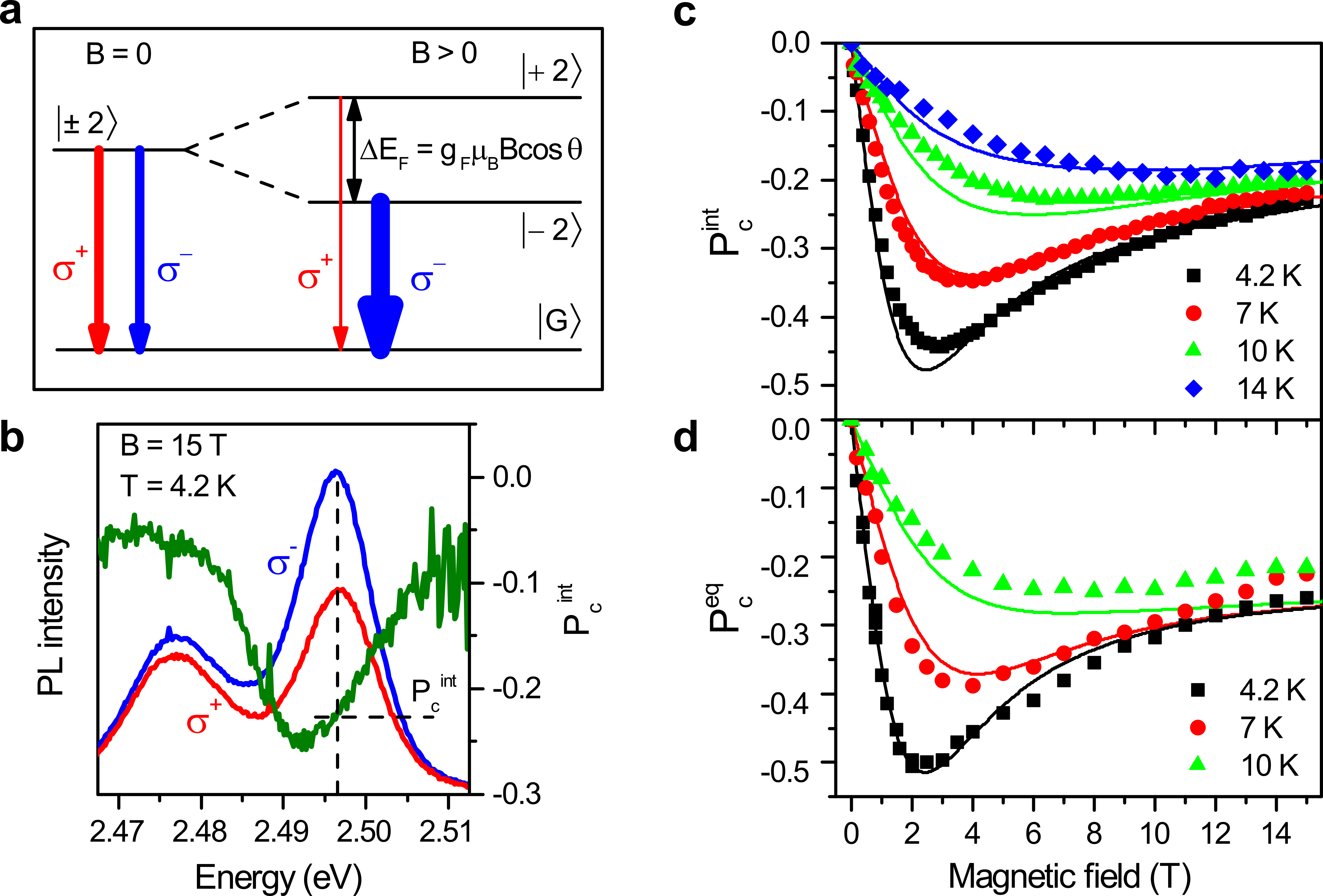}
	\caption{\textbf{Exciton polarization in Sample~1.} \textbf{a,} Scheme of the optical transitions between the dark exciton state $\ket{\pm 2}$, which is split in magnetic field into two spin sublevels ($g_{\rm F}>0$), and the unexcited ground state of the crystal $\ket{G}$.  At cryogenic temperatures, the lowest spin state $\ket{- 2}$ has higher occupation, and the emission is $\sigma^-$ circularly polarized. \textbf{b,} Time-integrated PL intensity of $\sigma^{-}$ (blue) and $\sigma^{+}$ (red) circularly polarized emission components in magnetic field of $B=15$~T. Time-integrated DCP (green) is calculated according to equation~\eqref{eq:DCP_definition}. $P_{\rm c}^{\rm int}$ at $2.497$~eV is marked by the horizontal dashed line. \textbf{c, d,} Magnetic field dependences of the time-integrated DCP, $P_{\rm c}^{\rm int}$, and  equilibrium  DCP, $P_{\rm c}^{\rm eq}$,  at various temperatures measured at the exciton PL maximum of $2.497$~eV. Symbols are experimental data, curves are calculations with parameters given in Table 1 and Methods. For modeling the contribution of the spin-dependent recombination is excluded ($\Gamma_{\rm db} = 0$).
	 } 
	\label{fig:Fig2abcd}
\end{figure*}

\textbf{Polarized PL in magnetic field.} Due to the large bright-dark exciton splitting, the PL at low temperatures is determined by the recombination of dark excitons. In applied magnetic field, the dark exciton spin sublevels  $\ket{\pm 2}$ split due to Zeeman effect by the energy $\Delta E_{\rm F}(B, \theta)=\Delta E_{\rm F}(B) \cos \theta=g_{\rm F}\mu_{\rm B} B \cos \theta$, where $g_{\rm F}$ is the dark exciton $g$ factor, $\mu_{\rm B}$ is the Bohr magneton and $\theta$ is the angle between the NPL quantization c-axis, which is normal to the NPL surface, and the direction of magnetic field \cite{Efros1996}. At low temperatures ($kT<\Delta E_{\rm F}(B)$ with $k$ being Boltzmann constant) the lower Zeeman sublevel is dominantly populated, and the emission is circularly polarized. The spin projection of the lowest Zeeman exciton sublevel determines the polarization sign (Fig.~\ref{fig:Fig2abcd}a). In Sample 1, in a magnetic field the lowest sublevel has spin projection $\ket{-2}$, so that the $\sigma^{-}$ polarized emission has higher intensity (Fig.~\ref{fig:Fig2abcd}b). 

The degree of circular polarization (DCP) of PL  is defined as
\begin{equation}
P_{\rm c} (t)=\frac{I^+(t) - I^-(t)}{I^+(t) + I^-(t)},
\label{eq:DCP_definition}
\end{equation}
at a specific moment of time $t$, where $I^+$ and $I^-$ are the intensities of $\sigma^+$ and $\sigma^-$ circularly polarized emission. The time-integrated and equilibrium DCP are denoted as $P_{\rm c}^{\rm int}$ and $P_{\rm c}^{\rm eq}$, respectively. As the Zeeman splitting $\Delta E_{\rm F}(B)$ varies with magnetic field, the population of the exciton Zeeman sublevels changes correspondingly to the Boltzmann distribution function. One expects that $P_{\rm c}^{\rm eq}$ increases with  magnetic field up to the saturation level $P_{\rm c}^{\rm sat}$, which depends on the NCs orientation in the ensemble:
\begin{equation}
P_{\rm c}^{\rm eq}(B)= P_{\rm c}^{\rm sat} \tanh \frac{\Delta E_{\rm F}(B)}{2kT}  .
\label{eq:Psat} 
\end{equation}
Such behavior is typically observed in colloidal NCs.\cite{JohnstonHalperin2001,Furis2005,Turyanska2010,Liu2013,Siebers2015}

The experimental values of $P_{\rm c}^{\rm int}$ and $P_{\rm c}^{\rm eq}$ measured for Sample~1 at the exciton maximum of $2.497$~eV are plotted in Fig.~\ref{fig:Fig2abcd}c,d. Note that the exciton spin relaxation time $\tau_{\rm s}<1$~ns (Supplementary Section~\ref{subsec:SI_TR(B)_Theor}) is considerably shorter than the exciton lifetime $\tau$ and the dynamical factor $\tau/(\tau+\tau_{\rm s})\approx 1$.\cite{Liu2013} Let us consider $P_{\rm c}^{\rm eq}(B)$ at $T=4.2$~K. It reaches a high value of $-50 \%$ at $B=2.5$~T  and then decreases to $-27 \%$ at 15~T. Two facts differ this dependence from our expectations according to equation \eqref{eq:Psat}. First, the very fast initial shift is described by $g_{\rm F}=5$ at $T=4.2$~K and  $g_{\rm F}=3.5$ at $T=7$~K if we assume $P_{\rm c}^{\rm sat}=-1$. 
This is considerably larger than $g_F=2$ in the studied NPLs (see Methods) and has unusual temperature dependence of $g$ factor. Second, instead of saturation with increasing magnetic field it strongly decreases. With increasing temperature the maximum becomes less pronounced and for $T> 10$~K the DCP turns into a monotonic increase with saturation. It is evident, that additional mechanisms controlling the exciton spin polarization are involved. The most reliable candidate is related to surfaces magnetism provided by the dangling bond spins, which has been demonstrated experimentally in CdSe NCs.\cite{Biadala2017} The physics here has close analogy with diluted magnetic semiconductors, where the carrier exchange with localized spins of magnetic Mn$^{2+}$ ions results in giant Zeeman splitting \cite{Furdyna1988}, which competes with the intrinsic splitting.\cite{Wojtowicz1999}

\begin{figure}[h!]
	\includegraphics{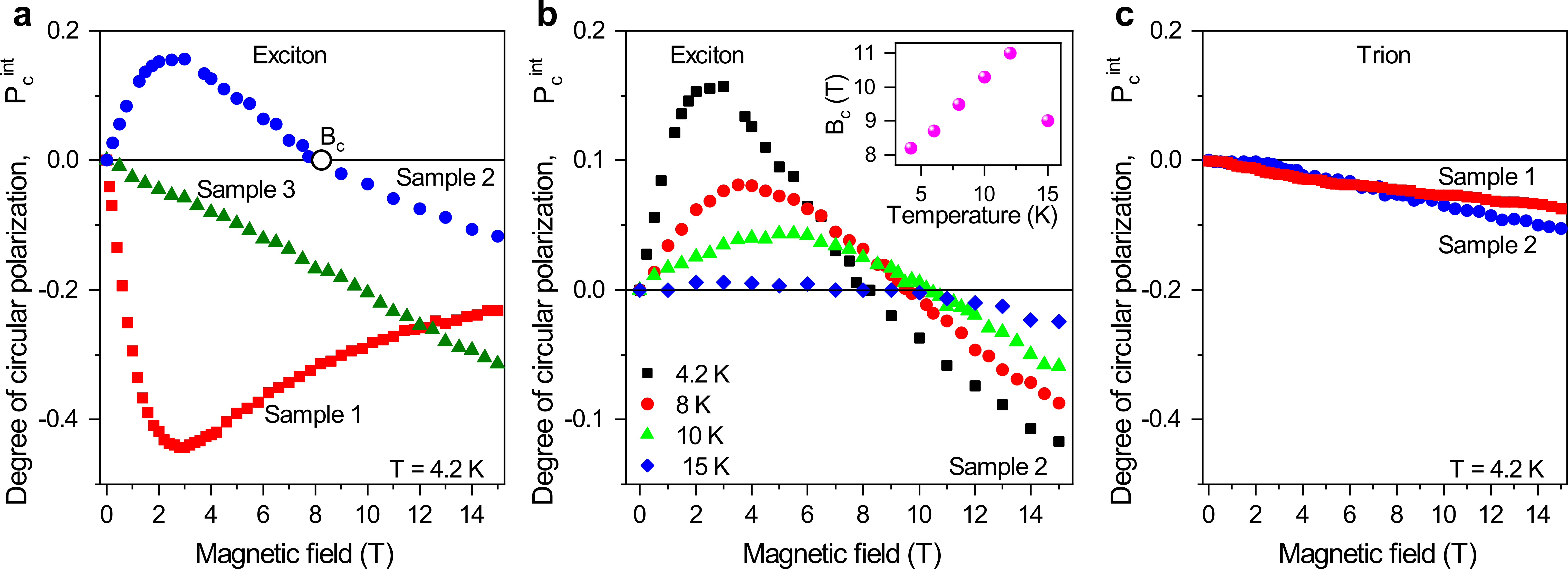}
	\caption{\textbf{Exciton and trion polarization in different samples.} \textbf{a,} Time-integrated DCP $P_{\rm c}^{\rm int}(B)$ of exciton in Samples~1 (red squares), 2 (blue circles), and 3 (green triangles). The critical magnetic field $B_{\rm c}$ for $P_{\rm c}^{\rm int}=0$ is marked by a circle. \textbf{b,} Exciton $P_{\rm c}^{\rm int}(B)$ at various temperatures in Sample~2. Inset: Critical magnetic field as a function of temperature. \textbf{c,} Time-integrated DCP of trion in Samples~1 (red) and 2 (blue). 
	}
	\label{fig:Pint(B)_3samples}
\end{figure}

A different, but also non-monotonous behavior is observed for Sample~2 (Fig.~\ref{fig:Pint(B)_3samples}a, blue): compared to Sample 1, the DCP has the opposite sign in low magnetic fields. It increases first to $+15 \%$ at $B=3$~T and then decreases with crossing zero polarization at  $B_{\rm c}=8.2$~T and reaching $-11\%$ at 15~T. With temperature increase, the DCP decreases and becomes close to zero at $T \ge 15$~K (Fig.~\ref{fig:Pint(B)_3samples}b).  The critical magnetic field $B_{\rm c}$ first shifts to higher values and decreases above $T=12$~K  (Fig.~\ref{fig:Pint(B)_3samples}b, inset). The non-monotonous DCP is not unique for the two samples shown here, but is typical for bare core CdSe NPLs (Supplementary Fig.~\ref{fig:SI_Fig5}b).

The polarization properties of the negatively charged excitons in Samples 1 and 2, which emission is shifted to lower energies by $20-26$~meV from the exciton line, are very different from the exciton. Trion DCP has a weak monotonic increase reaching $-7.5\%$ and $-10\%$ at $B=15$~T in Samples~1 and 2, respectively (Fig.~\ref{fig:Pint(B)_3samples}c). The negatively charged exciton consists of two electrons in a singlet state with zero total spin and a hole, which Zeeman splitting controls the trion DCP. The negative DCP is in line with trion properties in CdSe/CdS NPLs with thick shells.~\cite{Shornikova2018nl} The small and monotonous trion DCP let us conclude that the hole exchange interaction with surface spins is very weak. An obvious reason for that is small overlap of the hole wave function with the surface states, as hole is strongly localized in CdSe core due to its heavy mass. 

It is instructive to modify CdSe NPLs by ZnS shells. It strongly decreases the electron overlap with surface states, but also changes the surface itself, which is not Cd-terminated anymore. Indeed, DCP in Sample 3 of CdSe/ZnS NPLs differs from Samples 1 and 2. It shows linear increase reaching $-0.30\%$ at $B=15$~T (Fig.~\ref{fig:Pint(B)_3samples}a, green triangles and Supplementary Fig.~\ref{fig:SI_Fig5}c). Note that about 80\% of the emission from Sample 3 is contributed by excitons (Supplementary Section~\ref{subsec:SI_FStr_TR(T)_S3}). This comparison strongly supports our conclusion that in CdSe NPLs the electron-DBS exchange is responsible for the exciton interaction with surface spins.

\textbf{Theoretical consideration and discussion}

\begin{figure}[t]
	\includegraphics{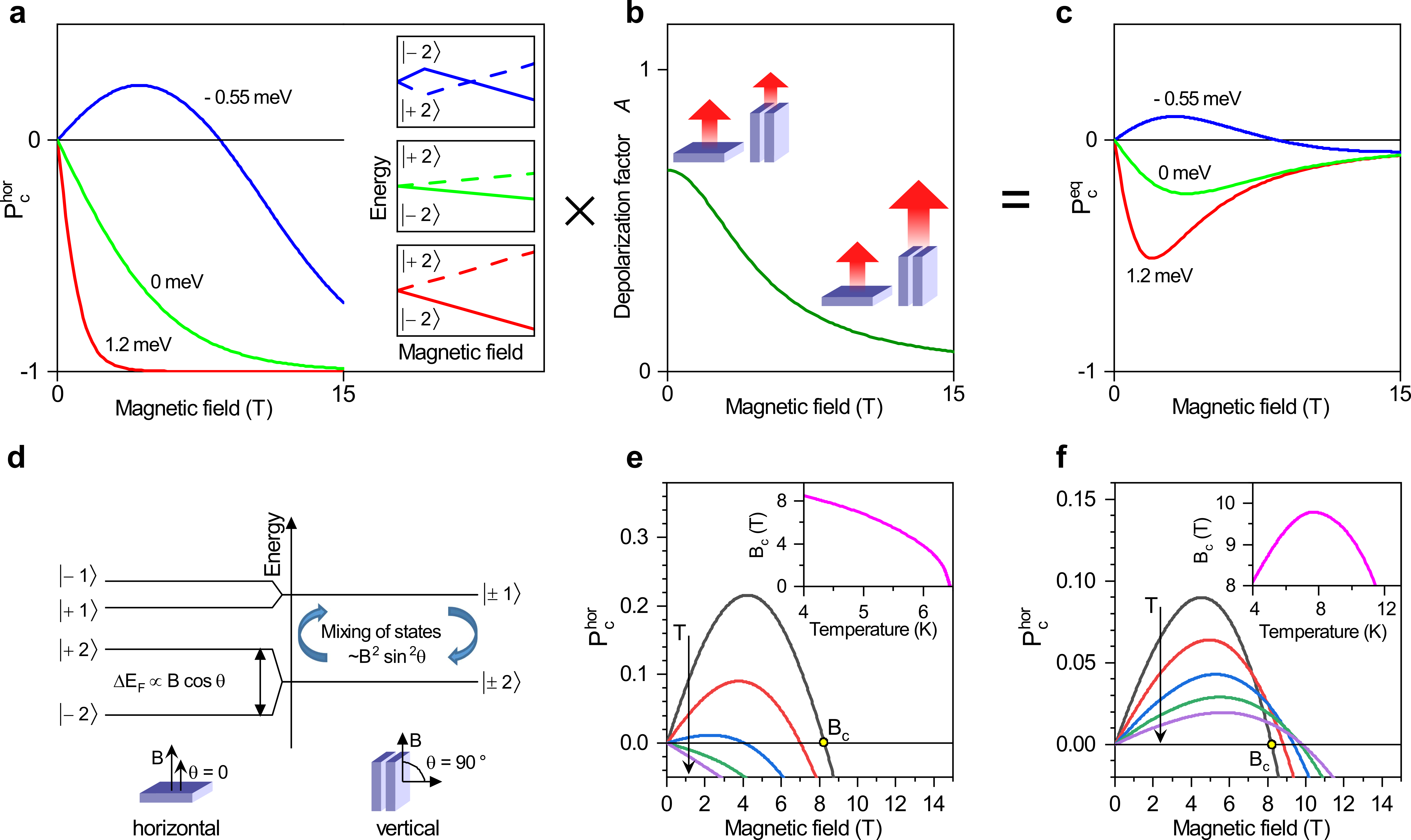}
	\caption{\textbf{Concept and theory.} \textbf{a.}  Modeled DCP of horizontally oriented NPLs  at $T=4.2$~K  with  $g_{\rm F}=2$ and different $E_{\rm p} = 1.2$~meV (red), $0$~meV (green), and $-0.55$~meV (blue). Insets show schematically corresponding exciton Zeeman splittings. \textbf{b.} Depolarization factor $A(B)=I_{\rm hor}/(I_{\rm hor}+I_{\rm ver})=2\Gamma_{\rm hor}/(2\Gamma_{\rm hor}+\eta\Gamma_{\rm ver})$ with $\eta=1$. \textbf{c.} Equilibrium DCP of NPL ensemble $P_{c}^{\rm eq}(B)=A(B) P_c^{\rm hor}(B)$.  \textbf{d.} Schematic presentation of exciton Zeeman sublevels in external magnetic field. The Zeeman splitting is large in horizontally oriented NPLs (left), and zero in vertically oriented ones (right). The emission of the latter is accelerated in magnetic field due to mixing with bright states. \textbf{e, f.} Magnetic field dependence of $P_{\rm c}^{\rm hor}(B)$ in ensemble of horizontal NPLs at temperatures of 4, 5, 6, 7, and 8~K (arrows show temperature increase),  $g_{\rm F}=2$. \textbf{e:} No spin-dependent recombination of the dark exciton ($\Gamma_{\rm db}/\Gamma_{\rm F}^0=0$), $E_{\rm p}=-0.55$ meV. \textbf{f:} Strong spin-dependent recombination ($\Gamma_{\rm db}/\Gamma_{\rm F}^0=0.91$),  $E_{\rm p}=-0.1$~meV. Insets show corresponding temperature dependence of critical magnetic field.}
	\label{fig:Fig4_theory_main}
\end{figure}

We turn to theoretical model of exchange interaction between the dark exciton and the surface spins to understand the drastically different DCP in Samples 1 and 2.  Illustrations in Fig.~\ref{fig:Fig4_theory_main} demonstrate the key features of the model, and comparison with experiment for Sample~1 is given in Fig.~\ref{fig:Fig2abcd}. 

{\bf Exciton Zeeman splitting.} According to equation  (\ref{eq:Psat}), DCP of the dark exciton reflects the exciton spin polarization caused by its Zeeman splitting. The exciton Zeeman splitting is contributed by an exchange field created by the surface spins polarized in  external magnetic field~\cite{Rodina2018JEM}: 
\begin{eqnarray}
\Delta  E_{\rm F}(B,\theta)=\left[ g_{\rm F}\mu_{\rm B}B +  2 E_{\rm p}  \rho_{\rm db}(B)\right] \cos \theta =  \Delta  E_{\rm F}(B)\cos \theta\, .
\label{Zdb}
\end{eqnarray}
Here $E_{\rm p}$ is the exciton-DBS exchange interaction energy, $\rho_{\rm db} = (n_{\rm db}^{-}-n_{\rm db}^{+})/(n_{\rm db}^{-}+n_{\rm db}^{+})$ is the spin polarization of the dangling bonds with
$n_{\rm db}^{+}$ ($n_{\rm db}^{-}$)  being the surface density of the dangling bonds with spin parallel (antiparallel) to the magnetic field direction.~\cite{Rodina2018JEM} Hereafter, we assume that $\rho_{\rm db}=\tanh ({g_{\rm db}\mu_{B}B}/{2kT})$ with $g$ factor $g_{\rm db}=2$ typical for surface paramagnetic centers in CdSe.\cite{Ditina1968}  In low magnetic fields ($g_{\rm db}\mu_{B}B < 2kT$) splitting is given by $
\Delta  E_{\rm F}(B) \approx (g_{\rm F}+E_{\rm p}g_{\rm db}/kT) \mu_{\rm B}B
$. If $g_{\rm F}$ and $E_{\rm p}g_{\rm db}$ have the same signs, the exchange field of surface spins increases the exciton Zeeman splitting. In case of opposite signs, it decreases the splitting and can even change its sign if $|g_{\rm F}|<|E_{\rm p}g_{\rm db}/kT|$. However, due to saturation of $\rho_{\rm db}$ with increasing field, the $\Delta E_{\rm F}(B)$ scales non-monotonically with $B$, crossing zero at $B_0=2|E_{\rm p}\rho_{\rm db}|/|g_{\rm F}|\mu_{\rm B}$ and changing sign (Supplementary Fig.~\ref{fig:dEf_SI}a).  Examples of $\Delta E_{\rm F}(B)$
dependencies modeled with Eq.~\eqref{Zdb} for different $E_{\rm p}$ are shown in Supplementary Fig.~\ref{fig:dEf_SI}a,b. 

\textbf{\bf DCP in horizontal NPLs.} 
The exciton Zeeman splitting  is maximal for the horizontally oriented NPLs for magnetic field applied perpendicular to the substrate, i.e. parallel to the c-axis (case $\theta=0$ in Fig.~\ref{fig:Fig4_theory_main}d). Exciton-DBS interaction    affects the DCP from horizontal NPLs,  $P_{\rm c}^{\rm hor}(B)=(I_{\rm hor}^{+}-I_{\rm  hor}^{-})/(I_{\rm  hor}^{+}+I_{\rm  hor}^{-})$ via its contribution to $\Delta E_{\rm F}(B)$. In addition, electron-DBS interaction   provides efficient mechanism of radiative recombination for the dark exciton by simultaneous flip of the electron spin in an exciton and a dangling bond spin.\cite{Rodina2015,Rodina2016} The mechanism efficiency depends on the number of DBSs oriented opposite to the electron spin. In zero magnetic field the DBSs have no preferable orientation and  the dark exciton states $\ket{\pm 2}$ have the same radiative recombination rate $\Gamma_{\rm hor}^{\pm 2}(0)=\Gamma_{\rm F}^0 =\Gamma_{0}+\Gamma_{\rm db}$. Here $\Gamma_{\rm db}$ is the DBS-assisted recombination rate and $\Gamma_{0}$ is the recombination rate acquired through other activation mechanisms, e.g. by interaction with phonons. In magnetic field the rates $\Gamma_{\rm hor}^{+2}$ and $\Gamma_{\rm hor}^{-2}$ become different:\cite{Rodina2018JEM}
$\Gamma_{\rm hor}^{\pm 2}(B)=\Gamma_{\rm F}(B) + \Gamma_{\rm db}[1 \pm \rho_{\rm db}(B)]=\Gamma_{\rm F}^0+\Gamma_{\rm F}^{\rm B} \pm \Gamma_{\rm db} \rho_{\rm db}(B)$, where  $\Gamma_{\rm F}$ ($\Gamma_{\rm F}^{\rm B}$) is the total (acquired) radiative recombination rate in magnetic field not related to DBSs. 
 
The resulting DCP of horizontal NPLs accounting for the spin-dependent recombination and the Zeeman splitting contributed by the DBSs reads (Supplementary Section~\ref{subsec:SI_Theor_DCP_hor})
\begin{eqnarray}
P_{\rm c}^{\rm hor}=\frac{-\rho_{\rm ex}(\Gamma_{\rm F}+\Gamma_{\rm db})+\rho_{\rm db}\Gamma_{\rm db}}
{\Gamma_{\rm F}+\Gamma_{\rm db}(1-\rho_{\rm ex}\rho_{\rm db})}  \, . \label{Phor}
\end{eqnarray}
Equation (\ref{Phor}) is valid for the time-dependent, time-integrated or equilibrium DCP, if one considers the time-dependent, time-integrated or equilibrium exciton polarization $\rho_{\rm ex}= (p_{\rm F}^{-}-p_{\rm F}^{+})/(p_{\rm F}^{-}+p_{\rm F}^{+})$, where $p_{\rm F}^{\pm}$ are occupation probabilities $p_{\rm F}^{\pm}$ of the dark exciton Zeeman sublevels. In case of the fast exciton spin relaxation, time-integrated $\rho_{\rm ex}$ coincides with  equilibrium value $\rho_{\rm ex}=\tanh(\Delta E_{\rm F}/2kT)$ even in the presence of the spin-dependent recombination (Supplementary Section~\ref{subsec:SI_Theor_DCP_hor}).   

In absence of the spin-dependent radiative recombination ($\Gamma_{\rm db}=0$), the DCP is proportional to the exciton polarization: $P_{\rm c}^{\rm hor}=-\rho_{\rm ex}$  contributed by the exciton-DBS exchange. Its magnetic field dependencies in horizontal NPLs are  shown in Fig.~\ref{fig:Fig4_theory_main}a for different exchange energies: $E_{\rm p}=1.2$, 0, and $-0.55$~meV. The corresponding ordering of exciton Zeeman sublevels are shown schematically in the inserts. For $E_{\rm p}=0$~meV (green) $\Delta E_{\rm F}(B)$ is just the intrinsic Zeeman splitting of the dark exciton with $g_{\rm F}=2$, which positive value results in a negative $P_{\rm c}^{\rm hor}$. $E_{\rm p}=1.2$~meV (red) corresponds to a ferromagnetic exchange of the dark exciton with the DBSs, which increases $\Delta E_{\rm F}(B)$ and, correspondingly, $P_{\rm c}^{\rm hor}$. For an antiferromagnetic exchange with $E_{\rm p}=-0.55$~meV (blue), in weak magnetic fields the exchange with DBSs is stronger than the intrinsic Zeeman splitting and the order of exciton spin levels is reversed giving rise to positive $P_{\rm c}^{\rm hor}$. However, the non-monotonic dependence of $\Delta E_{\rm F}(B)$ results into non-monotonic dependence of $P_{\rm c}^{\rm hor}(B)$ with changing sign at $B_{\rm c}=B_0$. 

The spin-dependent radiative recombination ($\Gamma_{\rm db}\neq 0$) affects the DCP  and for large values of $\Gamma_{\rm db}/\Gamma_{\rm F}$ equation (\ref{Phor}) leads to $P_{\rm c}^{\rm hor}=(\rho_{\rm db}-\rho_{\rm ex})/(1-\rho_{\rm ex}\rho_{\rm db})$. 
The  DCP of horizontal NPLs with account for $\Gamma_{\rm db} \ne 0$ according to  equation \eqref{Phor} are shown in Supplementary Figure~\ref{fig:SIEdbGdb}. The effect of $\Gamma_{\rm db} \ne 0$ on  $P_{\rm c}^{\rm hor}$ is dramatic for the case of small or negative values of $E_{\rm p}$, but very small  for  $E_{\rm p}=1.2$~meV. 

{\bf DCP in Sample 1.}
One can see in Fig.~\ref{fig:Fig4_theory_main}a that $P_{\rm c}^{\rm hor}$ at  $E_{\rm p}=1.2$~meV increases monotonically reaching saturation level $P_c^{\rm hor} \rightarrow -1$  at large magnetic fields. In experiment smaller and non-monotonous DCP is observed in Sample~1.  The DCP saturation can be reduced by account of the vertical NPLs with $\theta=\pi/2$ that may exist in stacks and  deliver unpolarized contribution to PL because of zero Zeeman splitting. The resulting DCP of the NPL ensemble is 
\begin{equation}
P_{\rm c}(B)=\frac{I_{\rm hor}(B) }{I_{\rm hor}(B)+I_{\rm ver}(B)} P_{\rm c}^{\rm hor}(B)= A(B)P_{\rm c}^{\rm hor}(B)   \, .
\label{Pgen}
\end{equation}
Here $I_{\rm hor}$($I_{\rm ver}$) is the complete PL intensity from horizontal (vertical) NPLs. The depolarization factor $A$ is responsible for the reduction of the DCP saturation value $P_{\rm c}^{\rm sat}$. It was found to be field independent for the trion emission in CdSe/CdS NPLs \cite{Shornikova2018nl}. For the dark exciton PL,  the magnetic field dependence of the depolarization factor  $A(B)= 2\Gamma_{\rm hor}(B)/[2\Gamma_{\rm hor}(B)+\eta \Gamma_{\rm ver}(B)]$  arises via different field dependencies of the exciton radiative recombination rates $\Gamma_{\rm hor}(B)$ and $\Gamma_{\rm ver}(B)$ in horizontal and vertical NPLs, while the lifetimes $\tau_{\rm L,M}$ are the same in both orientations (Methods and Supplementary Section~\ref{subsec:SI_TR(B)_Theor}). Here $\eta$ is  the ratio of photoexcited vertical and horizontal NPLs.

The field dependence of the  depolarization factor $A(B)$ is shown in Fig.~\ref{fig:Fig4_theory_main}b. It is calculated with $\Gamma_{\rm hor}(B)=\Gamma_0$ and  $\Gamma_{\rm ver}(B)=\Gamma_0+\Gamma_{\rm F}^B(B,\pi/2)$ from equation~\eqref{EQ:gammaFBM}  with parameters for Sample 1 from Table 1, $g_{\rm e}=1.7$ and  $\eta=1$, corresponding to equal concentrations of horizontal and vertical NPLs. The decrease of $A(B)$ with growing magnetic field is caused by the magnetic-field-induced increase of $\Gamma_{\rm ver}(B)$. The cumulative effect of $P_{\rm c}^{\rm hor}(B)$ and $A(B)$ for equilibrium polarization is shown in Fig.~\ref{fig:Fig4_theory_main}c. Note, that for  $P_{\rm c}^{\rm eq}(B)$ only NPLs with lifetime $\tau_{\rm L}$ should be considered. Important here is that for $E_{\rm p}>0$ the DCP becomes non-monotonic due to the decrease of $A(B)$. Also the absolute values are reduced compared to $P_{\rm c}^{\rm hor}(B)$ in Fig.~\ref{fig:Fig4_theory_main}a.  The ratio of vertical to horizontal NPLs, characterized by $\eta$, controls the maximum DCP value and the saturation value  $P_{\rm c}^{\rm sat}$  in high magnetic fields (Supplementary Fig.~\ref{fig:eta_SI}), which allows one to evaluate $\eta$ from the measured $P_{\rm c}^{\rm eq}(B)$.   

Equation~\eqref{Pgen} allowes us to model $P_{\rm c}^{\rm eq}(B)$ and  $P_{\rm c}^{\rm int}(B)$ in Sample~1 measured at different temperatures (Fig.~\ref{fig:Fig2abcd}c,d).  Details and parameters are given in Methods and Supplementary Sections~\ref{subsec:SI_Theor_Fitting_Peq}, \ref{subsec:SI_Theor_Fitting_Pint}.  At low temperatures the exciton-DBS interaction increases the  DCP slope in weak fields  and  the strongly non-monotonic behavior is obvious. It softens with increasing temperature due to decrease of the exciton ($\rho_{\rm ex}$) and DBS ($\rho_{\rm db}$) polarizations. In the modeling we use only few fitting parameters  and there is not much freedom of choice for their values as they control different features of the DCP dependence. In particular, the best fit is received with $\Gamma_{\rm db}=0$. We conclude that the DCP  in Sample~1  is dominated by the exciton-DBS exchange contribution to the  Zeeman splitting,  while the DBS-assisted exciton recombination is not important. 

{\bf Spin-dependent recombination in Sample~2.} 
Strongly non-monotonic and sign-reversal DCP dependence on magnetic field in Sample~2 (Fig.~\ref{fig:Pint(B)_3samples}a) is characterized by the critical field $B_{\rm c}=8.2$~T at $T=4.2$~K, where the exciton DCP vanishes. It can be modeled within two approaches distinguished by the role of spin-dependent recombination. The first one disregards it ($\Gamma_{\rm db}=0$) and requires $E_{\rm p}=-0.55$~meV (Fig.~\ref{fig:Fig4_theory_main}e). In this case at $B_{\rm c}=B_0=2|E_{\rm p}| \rho_{\rm db}/g_{\rm F}\mu_{\rm B}$, also $\Delta E_{\rm F}$=0, and, consequently, $\rho_{\rm ex}=0$. The exchange contribution decreases with increasing temperature due to depolarization of surface spins. As a result, $B_{\rm c}=B_0$ should shift to smaller values (Supplementary Fig.~\ref{fig:dEf_SI}d and Fig.~\ref{fig:Fig4_theory_main}e, inset). This is, however, in contradiction with the experimental appearances of Sample~2, shown in Fig.~\ref{fig:Pint(B)_3samples}b for $P_{\rm c}^{\rm int}$ and Supplementary Fig.~\ref{fig:SI_Fig5}a for  $P_{\rm c}^{\rm eq}$.
The second approach, shown in Fig.~\ref{fig:Fig4_theory_main}f, accounts for the finite spin-dependent recombination ($\Gamma_{\rm db}/\Gamma^0_{\rm F}=0.91$)  and requires a smaller $E_{\rm p}=-0.1$~meV.      
For $\Gamma_{\rm db} \neq 0$,  $B_{\rm c} \ne B_0$  and is no longer determined by the condition $\Delta E_{\rm F}=0$, but by $\rho_{\rm ex}=\rho_{\rm db}\Gamma_{\rm db}/(\Gamma_{\rm F}+\Gamma_{\rm db})$, as not only the exciton polarization, but also its spin-dependent recombination dynamics are involved.
Importantly, in this case the temperature dependence of $B_{\rm c}$ looks very different, shifting with increasing temperature first to higher fields, then turns around and decreases (Fig.~\ref{fig:Fig4_theory_main}f, insert). This is in good agreement with the experimental results from Fig.~\ref{fig:Pint(B)_3samples}b. Therefore, we conclude that in Sample~2 the DCP dependencies are controlled by the strong DBS-assisted spin-dependent recombination. 

It is worth noting, that in Sample~2 the positive sign of DCP in weak fields is provided by the spin-dependent recombination. In its absence, at $E_{\rm p}=-0.1$~meV the DCP should be negative, as $2|E_{\rm p}| \rho_{\rm db} < g_{\rm F} \mu_{\rm B} B$ and $\Delta E_{\rm F}>0$ at all magnetic fields and temperatures (Supplementary Fig.~\ref{fig:dEf_SI}a,b,c) with the order of levels the same as for $E_{\rm p}=0$ (Fig.~\ref{fig:Fig4_theory_main}a, inset).  As one can see in Supplementary Fig.~\ref{fig:SIEdbGdb}b, in case of a small exciton-DBSs exchange splitting ($2|E_{\rm p}| \rho_{\rm db} \ll g_{\rm F} \mu_{\rm B} B$) and $g_{\rm F}<g_{\rm db}$, the spin-dependent recombination determines the DCP sign. 

The CdSe NPLs in Samples~1 and 2 differ not only by the role of the spin-dependent recombination for their exciton polarization, but also by the type of exciton-DBS exchange, i.e. the sign of $E_{\rm p}$. It is ferromagnetic in Sample~1 synthesized in argon atmosphere, but antiferromagnetic in Sample~2 prepared in air. Obviously, different surface states are responsible for that, whose origin still has to be disclosed. Note, that two different types of surface paramagnetic centers were identified by electron paramagnetic resonance on CdSe micropowders prepared in vacuum and in air.\cite{Ditina1968} They were tentatively assigned to intrinsic vacancies of Cd or Se surface atoms and to absorbed oxygen or water. 


\textbf{Conclusions}	

We have demonstrated experimentally that colloidal nanoplatelets made of diamagnetic CdSe semiconductor have paramagnetic surface spins. They can be detected optically via the dark exciton photoluminescence. Being polarized in external magnetic field the surface spins provide additional Zeeman splitting of the exciton states, as well as contribute to exciton spin-dependent radiative recombination. We have developed a theoretical formalism to account for these two factors in the circular polarization degree of exciton emission in magnetic field. Modeling of the data has allowed us to estimate the values of the exchange constants and exciton recombination rates and to establish their dependence on the synthesis conditions of the CdSe nanoplatelets. Their preparation in argon or air atmosphere results in different types of surface states interacting either ferro- or antiferromagnetically with the dark excitons. In the latter case the spin-dependent recombination is crucial. The comparison with the charged exciton in bare CdSe nanoplatelets and with the exciton in core/shell CdSe/ZnS nanoplatelets evidences that the electron interaction with the surface spins is the key factor for the observed effects. The demonstrated experimental approach to surface magnetism can be extended to study chemically or physically modified nanocrystals, e.g., by their interaction with magnetic substrates or magnetic nanocrystals. 
Spin functionalization of surfaces in colloidal structures is important for their use as nanomagnetic markers. Another promising application field may arise from the combination of surface magnetism with chiral organic ligands.

\subsection*{Methods}

\textbf{Samples.} 
We studied three batches of CdSe-based NPLs: (i) Sample 1 contains bare core CdSe NPLs with 4 monolayers thickness being synthesized in argon atmosphere, (ii) Sample 2 consists of bare core CdSe NPLs, also with 4 ML thickness synthesized in ambient conditions, and (iii) Sample 3 contains core/shell CdSe/ZnS NPLs with the same core thickness surrounded by 5-ML-thick ZnS shells from both sides. The sample parameters are given in Table~\ref{tab:table1}.
Sample~1 was characterized comprehensively in ref.~\onlinecite{Shornikova2018ns}, additional data for Samples~2 and 3 (TEM images, room temperature absorption and emission spectra, recombination dynamics at low temperatures) are given in the Supplementary Section~\ref{sec:Additional_S1_S2}. For CdSe NPLs most of the surface atoms are located in the planar surfaces, which are $[100]$-oriented with Cd-terminated facets,\cite{Li2011jacs} only a small fraction is on the side facets of 1~nm width, \textit{i.e.} most of surface atoms are Cd. 
Additionally, results for two further CdSe NPLs with 5 ML (Sample 4, in argon) and 4 ML (Sample 5, in air) are given in Supporting Information.   

Samples~1 and 4 were synthesized according to the protocol reported in ref.~\onlinecite{Ithurria2008} in argon atmosphere. Samples~2 and 5 were synthesized in ambient conditions and differ by lateral dimensions.
In a typical synthesis conducted under air, 70~mg of cadmium myristate 
and 24~mg of elemental selenium were added to a vial containing 7~mL of 
ODE. This mixture was heated, under stirring on a hot plate, to 210$^\circ$C 
and kept at this temperature for 10 minutes. Next, $90-180$~mg of cadmium 
acetate was added, and the mixture was kept at 210$^\circ$C for another 8~min. Finally, the mixture was cooled to 160$^\circ$C using an air flow, 1~mL of oleic acid was injected, and the vial was further cooled to room 
temperature. CdSe nanoplatelets were separated from the synthesis 
byproducts by selective precipitation and centrifugation, and finally 
suspended in hexane. All CdSe NPLs were passivated by oleic acid. 

\textbf{Optical measurements.}   
Concentrated NPL solutions were drop-casted onto a quartz plate and mounted on top of a three axis piezo-positioner in the variable temperature insert ($2.2-70$~K) of a liquid helium bath cryostat. External magnetic fields up to 15~T, generated by a superconducting solenoid, were applied in the Faraday geometry, i.e. parallel to the light propagation direction.
The PL was excited by a laser diode (405~nm, 3.06~eV) in cw or pulsed mode (pulse duration 50~ps, repetition rate varied between 0.8 and 5~MHz). The laser beam was focused into a spot with 200~$\mu$m diameter. The signal was collected through the same lens, dispersed by a 0.55-m monochromator with 600 or 1800 grooves/mm, and detected either by a liquid-nitrogen-cooled charge-coupled-device (CCD) camera or by an Si avalanche photodiode connected to a conventional time-correlated single-photon counting setup. The temporal resolution of the time-resolved measurements was 0.3~ns. For polarization-resolved measurements the PL was analyzed by a combination of a quarter-wave plate and a linear polarizer.

\textbf{Modeling details.}

{\bf a. Exciton  recombination dynamics in magnetic field.} 
The observed magnetic field dependencies of the dark exciton lifetimes $\tau_{\rm M}$ and $\tau_{\rm L}$  in the Sample 1, see the insert in Fig.~\ref{fig:Fig1ab}b,  are  described by  $1/\tau_{\rm L}=\Gamma_{\rm F}^{0}+\Gamma_{\rm F}^{\rm B}(B,\theta)$ and  $1/\tau_{\rm M} = 1/\tau_{\rm L}+\Gamma_{\rm nr}$, where $\Gamma_{\rm F}^{\rm B}(B,\theta)$ is the radiative rate acquired in a magnetic field not parallel to the c-axis, e.g., for the vertical NPLs (case $\theta=\pi/2$ in Fig.~\ref{fig:Fig4_theory_main}d), via admixture of the bright exciton state: \cite{Efros1996,Rodina2016,Rodina2018FTT}
\begin{equation}
\Gamma_{\rm F}^{\rm B}(B,\theta)= \left(\frac{g_e\mu_{\rm B}B\sin\theta}{2\Delta E_{\rm AF}}\right)^2\Gamma_{\rm A} \, .
\label{EQ:gammaFBM}
\end{equation}
Here $\Gamma_{\rm A}$ is the radiative recombination rate of the bright exciton, $g_{\rm e}$ is the in-plane electron $g$ factor. The parameters used are: $\theta=\pi/2$, $\Gamma_{\rm F}^{0}=0.009$ ns$^{-1}$, $\Gamma_{\rm nr} =  0.083$ ns$^{-1}$, $\Delta E_{\rm AF}=4.6$~meV, $\Gamma_{\rm A}=10~\rm{ns}^{-1}$ (Table~\ref{tab:table1}) and $g_{\rm e}=1.7$ from ref.~\onlinecite{Kalitukha2018}. The details and data for other samples are discussed in the Supplementary Section~\ref{subsec:SI_TR(B)_Theor}. 
We assume that in zero magnetic field  $\tau_{\rm L}$ and $\tau_{\rm M}$ correspond to two NPL subensembles, where recombination of the dark exciton is either purely radiative or has both radiative and nonradiative channels.  The subensemble with $\tau_{\rm M}$ can be associated with NPLs in stacks: It was reported that NPL stacking results in a 10-fold decrease of quantum yield compared to isolated NPLs.\cite{Guzelturk2014} Among possible reasons for the decrease is the energy transfer between NPLs in stacks delivering excitons to NPLs with nonradiative centers, as the inter-platelet distance is only 5~nm.\cite{Tessier2013acs} Transfer times of $6-10$~ps were reported.\cite{Rowland2015} The nonradiative centers can be deep trap states,\cite{Tessier2012,Kunneman2014,Guzelturk2014,Olutas2015} surface imperfections, etc.  The relative PL intensity of the middle component, $I_{\rm M}/(I_{\rm M}+I_{\rm L})$, increases as $B^2$ (Supplementary Section~\ref{subsec:SI_TR(B)_Theor}), indicating an increase of the radiative quantum yield of dark exciton in the subensemble with the nonradiative decay channel.

 Typically, the drop-casted NPLs have two preferable orientations on the substrate: vertical NPLs with c-axis parallel to the substrate plane, which are known to exist in stacks, and horizontal NPLs with c-axis perpendicular to the substrate plane with the magnetic field applied perpendicular to the substrate. These two orientations correspond to angles $\theta=\pi/2$ and 0, respectively.

 For the horizontally oriented NPLs, $\Gamma_{\rm F}^{\rm B}(B,0)=0$ [equation~\eqref{EQ:gammaFBM}] and  $\Gamma_{\rm F}(B)=\Gamma_0$. Even for tilted NPLs with $\theta=\pi/6$, the decay time $\tau_{\rm L}$ would exceed 25 ns in $B=15$~T. The experimentally observed PL decay (Fig.~\ref{fig:Fig1ab}b) does not contain such a long component at $B=15$~T. This may indicate a strong coupling or energy transfer between the horizontally and vertically oriented NPLs, resulting in the same life time for both orientations. An efficient F{\"o}rster energy transfer was indeed reported in NPLs.\cite{Guzelturk2014,Rowland2015} An analysis of the mechanisms resulting in the same magnetic field dependencies of the exciton life times in vertical and horizontal NPLs in our ensembles is beyond the scope of this article and will be presented elsewhere.  For further modeling of the magnetic field dependencies of $I_{\rm hor}(B)$ and $I_{\rm ver}(B)$ used in equation (\ref{Pgen}) and, thus, of the depolarization factor $A(B)$, we simply assume the same exciton lifetimes in horizontal and vertical NPLs.

With this assumption, the resulting  DCPs, $P_{c,{\rm M}} (B)$ and $P_{c,{\rm L}} (B)$,  in the subensembles with lifetimes $\tau_{\rm M}$ and $\tau_{\rm L}$, respectively, can be found as (see Supplementary Sections~\ref{subsec:SI_Theor_Intensities_Rates}, \ref{subsec:SI_Theor_Fitting_Peq}, \ref{subsec:SI_Theor_Fitting_Pint})
\begin{gather} \label{PgenG}
P_{\rm c,(L,M)}(B)=\frac{2\Gamma_{\rm hor}(B)}{2\Gamma_{\rm hor}(B)+\eta_{\rm L,M}\Gamma_{\rm ver}(B)}  P_{c,{\rm (L,M)}}^{\rm hor}(B) = A_{\rm (L,M)}  P_{c,{\rm (L,M)}}^{\rm hor}(B) \, .
\end{gather}
Here $\eta_{\rm L}$ and  $\eta_{\rm M}$ are the ratios of photoexcited vertical and horizontal NPLs with long lifetimes $\tau_{\rm M}$ and $\tau_{\rm L}$, respectively. 

\textbf{b. Equilibrium and time-integrated DCP.}
For the modeling and fitting of the DCP we take exciton $g$ factor $g_{\rm F}=g_{\rm e}-3g_{\rm h}\approx2$, being composed of the electron and hole $g$ factors: $g_{\rm e}\approx1.7$\cite{Kalitukha2018} and $g_{\rm h}\approx-0.1$. This hole $g$ value is in agreement with the trion DCP in Sample 1 (Supplementary Section~\ref{sec:SI_Suppl_DCP}) and with the theoretical value \cite{Shornikova2018nl}.  For the depolarization factor we use parameters for sample 1 given in the Table 1. 

The equilibrium DCP, $P_{c}^{\rm eq}$, corresponds to the DCP saturation at times $t \gg \tau_{\rm M}$. Therefore, it is contributed by  NPLs  with $\tau_{\rm L}$ only and $P_{c}^{\rm eq}(B)=P_{ c,{\rm L}}(B)$. 
To obtain a better fit of the data for $P_{\rm c}^{\rm eq}(B)$ in Fig.~\ref{fig:Fig2abcd}d , we additionally allow a weak field dependence of the spin-independent radiative rate in the horizontal NPLs  $\Gamma_{\rm F}(B)=\Gamma_0+\gamma \Gamma_{\rm F}^{\rm B}(B,\pi/2)$, where  $\gamma$ is a fitting parameter. Such dependence can be caused by the coupling to the excitons in vertical NPLs, by a small tilt angle for the horizontal NPLs.  The best fit shown by lines in Fig.~\ref{fig:Fig2abcd}d is achieved with the parameters used also for Figs.~\ref{fig:Fig4_theory_main}a-c, where $E_{\rm p}=1.2$~meV and $\gamma=0.16$. The use of $\gamma \ne 0$ softens the decrease of the depolarization factor with $B$ (Supplementary Fig.~\ref{fig:def_SI}), allowing for a larger saturation value of the DCP.

The best fit of the DCP data in Sample~1 shown in Fig.~\ref{fig:Fig2abcd}d is obtained without accounting for the spin-dependent recombination ($\Gamma_{\rm db}=0$). When the DBS-assisted spin-dependent recombination ($\Gamma_{\rm db}\ne 0$), the total radiative recombination rate of the dark exciton in horizontal NPLs  is given by $\Gamma_{\rm hor}(B)=\Gamma_{\rm F}(B)+\Gamma_{\rm db}[1-\rho_{\rm ex}(B)\rho_{\rm db}(B)]$. In the case $\rho_{\rm ex}(B)\rho_{\rm db}(B)>0$ this results in the 
prolongation of the exciton life time in magnetic field instead of shortenning.   The effect of the spin-dependent radiative recombination  on the exciton life time is similar to the effect of its nonradiative counterpart considered in the pioneering paper on spin-dependent recombination on silicon surface by Lepine.\cite{Lepine1972}  Note, that the spin-dependent nonradiative recombination affects the exciton lifetime and its polarization $\rho_{\rm ex}$,\cite{Paget1984} but maintains the direct proportionality between DCP and $\rho_{\rm ex}$.\cite{Weisbush1974} In contrast, the spin-dependent radiative recombination additionally affects the DCP (see equation \eqref{Phor} and  Supplementary Section~\ref{subsec:SI_Theor_DCP_hor}). 

Model calculations of the $P_{c}^{\rm eq}$ in sample 1 with  $\Gamma_{\rm db}/\Gamma_{\rm F}^0=0.5$  and $\Gamma_{\rm db}/\Gamma_{\rm F}^0 \approx 1$ are less reliable (Supplementary Fig.~\ref{fig:SIsample1Gdb}). As shown in Supplementary Figure~\ref{fig:SIEdbGdb}a, the effect of $\Gamma_{\rm db}/\Gamma_{\rm F}^0 \ne 0$ on  $P_{\rm c}^{\rm hor}$ for  $E_{\rm p}=1.2$~meV is very small. However, $\Gamma_{\rm db}/\Gamma_{\rm F}^0 \ne 0$  in this case sharpens the decrease of the depolarization factor with magnetic field. 

For the time-integrated DCP, the contributions of the two subensembles with long and middle lifetimes result in (Supplementary Section \ref{subsec:SI_Theor_Fitting_Pint})
\begin{gather}\label{Pint}
P_{\rm c}^{\rm int}(B)=\frac{P_{c,{\rm L}} (B)+P_{c,{\rm M}}(B)I_{\rm M}/I_{\rm L}}{1+I_{\rm M}/I_{\rm L}}, 
\end{gather}
where $I_{\rm M}/I_{\rm L}=A_{\rm M}\tau_{\rm M}/A_{\rm L}\tau_{\rm L}$ is the ratio of time-integrated intensities of the subensembles with 
$\tau_{\rm M}$ and $\tau_{\rm L}$ (see Supplementary Section~\ref{subsec:SI_TR(B)_Theor}). It can be seen from equations~\eqref{PgenG}--\eqref{Pint}, that the difference between $P_{\rm c}^{\rm eq}(B)$ and $P_{\rm c}^{\rm int}(B) <P_{\rm c}^{\rm eq}(B)$ is caused by the relation $P_{c,{\rm M}}<P_{c,{\rm L}}$ that can be provided by two factors: (i) a smaller DCP $P^{\rm hor}_{c,{\rm M}} < P^{\rm hor}_{c,{\rm L}}$ in horizontal NPLs and (ii) a larger number of vertical NPLs in the subensemble with lifetime $\tau_{\rm M}$ ($\eta_{\rm M}>\eta_{\rm L}$). Fitting of the experimental dependencies for $P_{\rm c}^{\rm int}(B )$ as presented in Fig.~\ref{fig:Fig2abcd}c was performed with $\eta_{\rm M}=1.3$,$\eta_{\rm L}=1$, $g_{\rm F}=2$, $\Gamma_{\rm db}=0$, $\gamma_{\rm M}=\gamma_{\rm L}=0.16$, $E_{\rm p}=1.2$~meV for the subensemble with $\tau_{\rm L}$ and $E_{\rm p}=0$ meV for the subensemble with $\tau_{\rm M}$.

\textbf{Data availability.} 
The data that support the plots within this paper and other findings of this study are available from the corresponding author upon reasonable request.

\subsection*{References}

\normalem 

\renewcommand{\refname}{References}

\subsection*{Acknowledgements}
The authors are thankful to Al. L. Efros and Yu. G. Kusrayev for fruitful discussions. We acknowledge the financial support by the Deutsche Forschungsgemeinschaft through the International Collaborative Research Centre TRR160 (Project B1),  the Russian Foundation for Basic Research (Grant No. 19-52-12064 NNIO-a). A.V.R. acknowledges partial support of the Russian Foundation for Basic Research (Grant No. 17-02-01063). A.P. and I.M. acknowledge funding from the European Research Council (ERC) under the European Union’s Horizon 2020 research and innovation programme (grant agreement no. 714876 PHOCONA).

\subsection*{Author  contributions}  
E.V.S. and A.A.G. contributed equally to this work.
E.V.S., L.B. and G.Q. performed the measurements under the guidance of D.Y. and M.B. A.A.G. and A.V.R. developed the theoretical model.  E.V.S., A.A.G., A.V.R. and D.R.Y. analyzed  and  interpreted  the  data. A.K. and M.N. synthesized nanocrystals (Samples 1, 3 and 4) under the guidance of B.D., and A.P. synthesized nanocrystals (Samples 2 and 5) under the guidance of I.M. E.V.S., A.A.G., A.V.R. and D.R.Y. wrote the manuscript with the assistance of all other co-authors. 

\subsection*{Additional information}
Supplementary  information  is  available in the online  version  of  the  paper. Reprints and permissions information is available online at www.nature.com/reprints. Correspondence  and  requests  for  materials  should  be  addressed  to  E.V.S. (elena.shornikova@tu-dortmund.de),  D.R.Y. (dmitri.yakovlev@tu-dortmund.de), and A.V.R. (anna.rodina@mail.ioffe.ru).

\subsection*{Competing  interests}
The  authors  declare  no  competing   interests.

\newpage

\clearpage
\widetext
\begin{center}
	\textbf{\large Supplementary Information:}
	
\vspace{3mm}	
	\textbf{\large Surface spin magnetism controls the polarized exciton emission from CdSe nanoplatelets}
	
\vspace{3mm}

Elena V. Shornikova,$^{1}$ Aleksandr A. Golovatenko,$^2$ Dmitri R. Yakovlev,$^{1,2}$ Louis Biadala,$^3$ Anna V. Rodina,$^2$ Gang Qiang,$^1$ Alexis Kuntzmann,$^4$ Michel Nasilowski,$^4$ Benoit Dubertret,$^4$ Anatolii Polovitsyn,$^{5,6}$ Iwan Moreels,$^{5,6}$ and Manfred Bayer$^{1,2}$
\end{center}
\vspace{3mm}

{\small \noindent$^1$Experimentelle Physik 2, Technische Universit{\"a}t Dortmund, 44221 Dortmund, Germany
	
\noindent$^2$Ioffe Institute, Russian Academy of Sciences, 194021 St. Petersburg, Russia

\noindent$^3$Institut d'Electronique, de Micro{\'e}lectronique et de Nanotechnologie, CNRS, 59652 Villeneuve-d'Ascq, France

\noindent$^4$Laboratoire de Physique et d'Etude des Mat\'{e}riaux, ESPCI, CNRS, 75231 Paris, France

\noindent$^5$Department of Chemistry, Ghent University, 9000 Ghent, Belgium}

\noindent$^6$Istituto Italiano di Tecnologia, 16163 Genova, Italy

\setcounter{equation}{0}
\setcounter{figure}{0}
\setcounter{table}{0}
\setcounter{section}{0}
\setcounter{page}{1}
\renewcommand{\theequation}{S\arabic{equation}}
\renewcommand{\thefigure}{S\arabic{figure}}
\renewcommand{\thetable}{S\arabic{table}}
\renewcommand{\thesection}{S\arabic{section}}

\section{Additional data for Samples 2 and 3:  TEM, absorption and emission spectra \label{sec:Additional_S1_S2}} \label{sec:SI_S1}

\begin{figure}[h!]
	\includegraphics[width=0.8\linewidth]{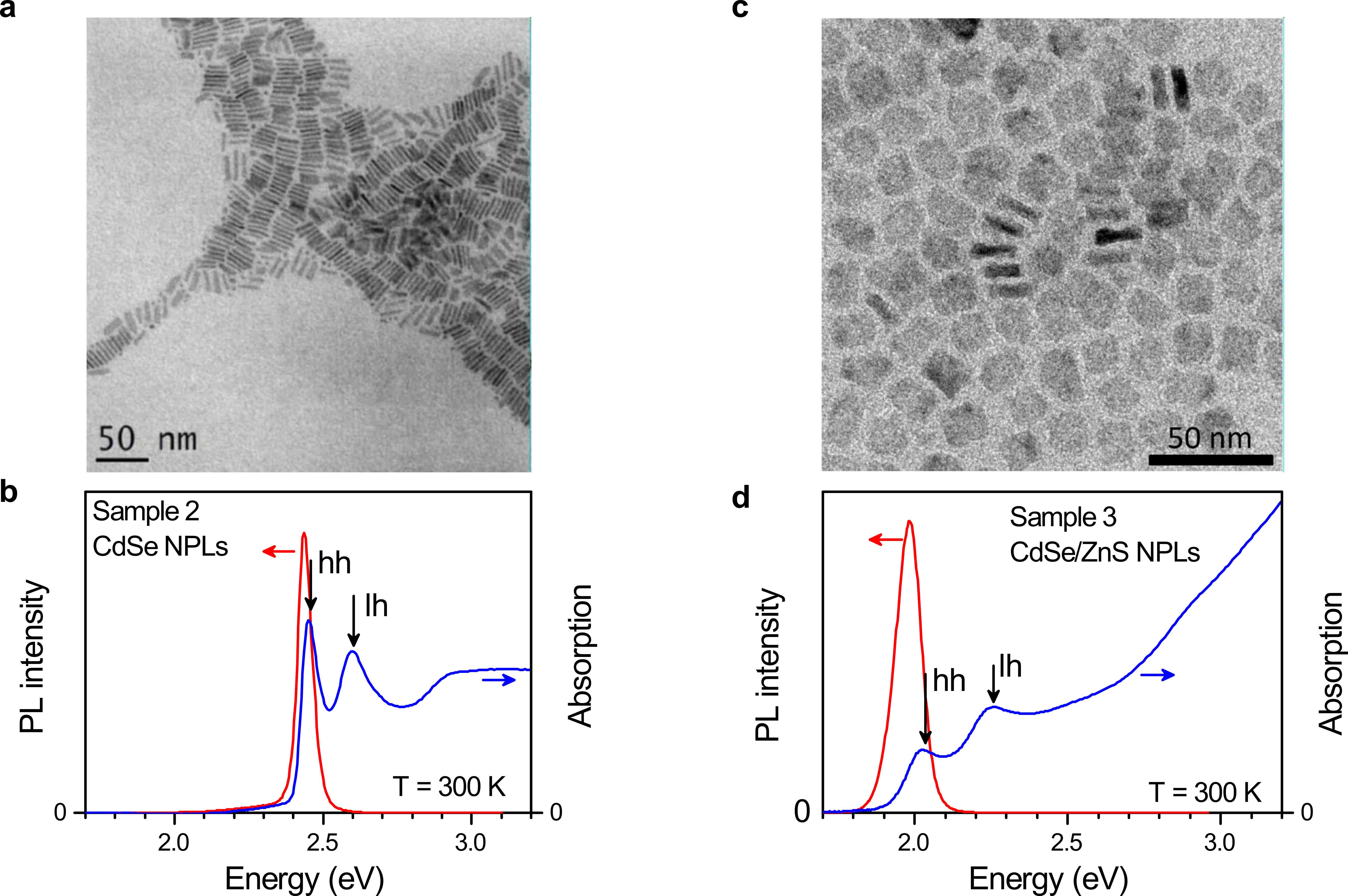}
	\caption{Characterization of Samples~2 and~3. Top: Transmission electron microscopy (TEM) images. Bottom: Room temperature photoluminescence (red) and absorption (blue) spectra. Heavy-hole (hh) and light-hole (lh) exciton absorption peaks are marked by black arrows. \textbf{a, b}, 1.2-nm-thick CdSe NPLs (Sample 2). \textbf{c, d}, core/shell CdSe/ZnS NPLs (Sample 3).}
	\label{fig:TEM}
\end{figure}

Transmission electron microscopy (TEM) images of samples 2 and 3 are shown in Figs.~\ref{fig:TEM}a and \ref{fig:TEM}c, respectively. The CdSe NPLs of Sample~2 have lateral dimensions of $4.9 \times 24$~nm$^2$, i.e. carriers are weakly confined in the smaller lateral direction. This partially explains the blue shift of the emission line compared to the one in Sample~1 having lateral dimensions of $8 \times 16$~nm$^2$. One can see in Fig.~\ref{fig:TEM}a that the CdSe NPLs of Sample 2 have two preferential orientation: either vertical (in stacks) or horizontal. 
The CdSe/ZnS NPLs of Sample~3 have lateral dimensions of about $18 \times 18$~nm$^2$ and the total thickness of 14 monolayers (MLs), being composed of 4 ML CdSe core and 5 ML ZnS shells on both sides. In the center of Fig.~\ref{fig:TEM}c one can see thin dark stripes corresponding to CdSe/ZnS NPLs vertically standing on their edges. These core/shell NPLs avoid stacking and tend to lay horizontally.

Room temperature absorption and emission spectra of bare CdSe NPLs (Sample~2) and core-shell CdSe/ZnS NPLs (Sample~3) are shown in Figs.~\ref{fig:TEM}b and ~\ref{fig:TEM}d, respectively. In the absorption spectra of both samples, two peaks corresponding to the heavy-hole and light-hole excitons are clearly seen. They are separated from each other by 146 and 220 meV in Samples 2 and 3, respectively. The full width at half maximum (FWHM) of the emission line in Sample 2 is 57~meV, and its maximum has a Stokes shift of 17~meV from the absorption peak. In Sample 3 these values are 98~meV and 40~meV, respectively. It is a characteristic feature of NPLs that the addition of the shell results in a broadening of the exciton lines.

\section{Exciton recombination dynamics and bright-dark energy splitting} \label{sec:SI_FStr}

\subsection{Three-level model} 
\label{subsec:SI_FStr_3-level}

\begin{figure}[h!]	
	\includegraphics{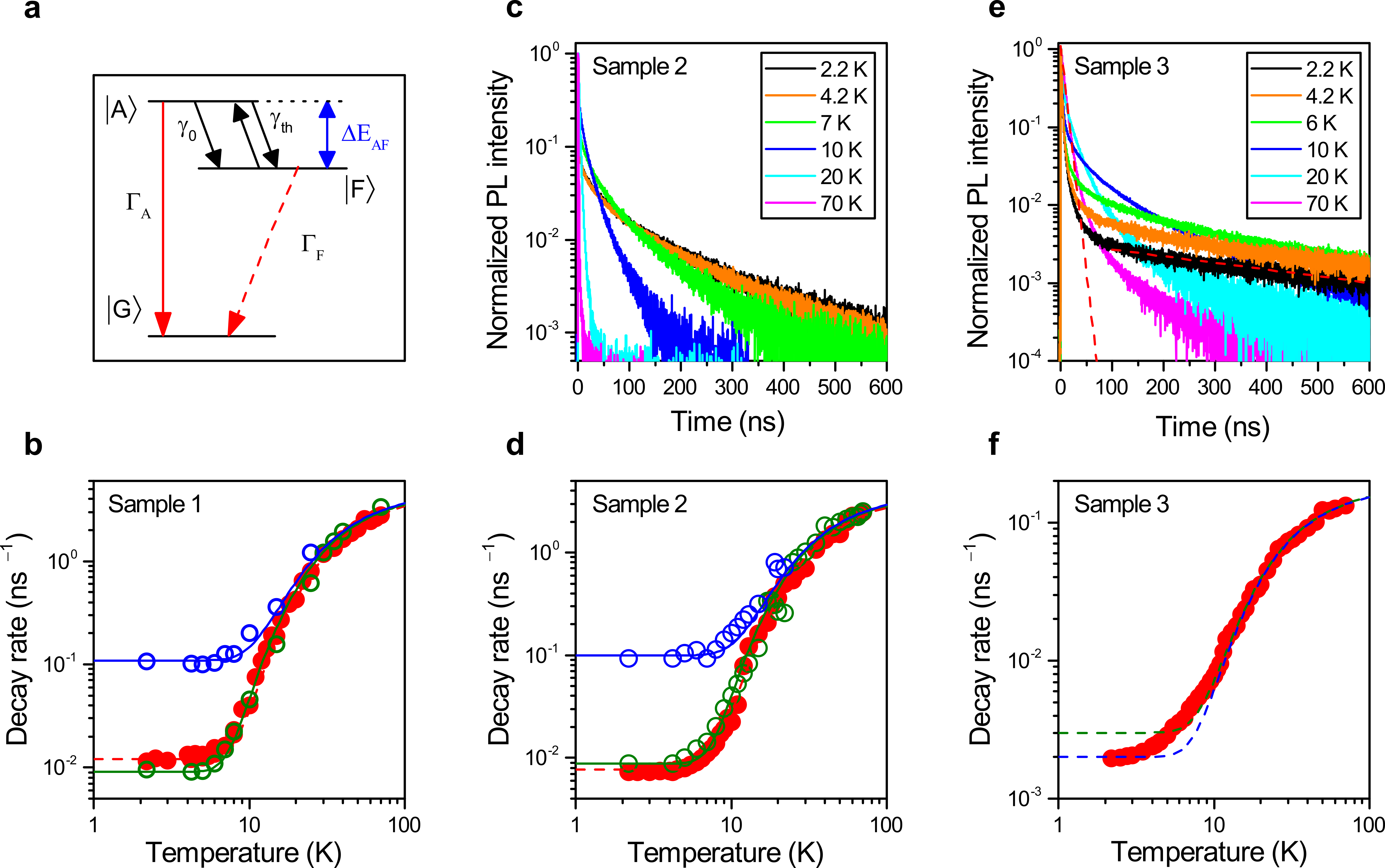}
	\caption{\textbf{Photoluminescence decay \textit{vs} temperature.} \textbf{a,} Three-level model: $\ket{A}$ and $\ket{F}$ are bright and dark exciton states, and $\ket{G}$ is unexcited crystal state. More details are given in ref.~\onlinecite{Shornikova2018ns}. \textbf{b,} Temperature dependence of the long component decay rate $\Gamma_{\rm L}$ for Sample 1, obtained from two-exponential fit (red symbols) and three-exponential fit (green symbols) of the PL decay. Blue symbols correspond to middle decay rate from three-exponential fit. Blue curve describes decay rate $\Gamma_{\rm M}=\Gamma_{\rm L}+\Gamma_{\rm nr}$ in the subensemble with a nonradiative recombination channel. The curves are fits with $\Delta E_{\rm AF}=5.0$~meV (red) and $\Delta E_{\rm AF}=4.6$~meV (green and blue). Recombination dynamics at different temperatures for Sample~1 can be found in Fig.~4a of ref.~\onlinecite{Shornikova2018ns}. \textbf{c,} PL decays at the maxima of exciton line of Sample~2 at various temperatures. \textbf{d,} Long and middle component decay rates $\Gamma_{\rm L}$ (green) and $\Gamma_{\rm M}$ (blue) in Sample~2 as function of temperature. Long decay component of two-exponential fit is presented by red circles. Lines are fits with equation (2) in ref.~\onlinecite{Shornikova2018ns}. The resulting $\Gamma_{\rm A}$, $\tau_{\rm L}=\Gamma_{\rm F}^{-1}$ and $\Delta E_{\rm AF}$ are given in Table~\ref{tab:table1}. \textbf{e,} PL decays at maxima of exciton line of Sample~3 measured at various temperatures. \textbf{f,} Long component decay rate $\Gamma_{\rm L}$ in Sample~3 as function of temperature. Lines are fits with equation (2) in ref.~\onlinecite{Shornikova2018ns} with $\Gamma_{\rm A} = 0.39$~ns$^{-1}$, $\Delta E_{\rm AF}=3.9$~meV, and $\Gamma_{\rm F} = 0.003$ (green) and $0.002$ (blue)~ns$^{-1}$. Since $\gamma_0$ is not known for Samples~2 and ~3, we used for their fits the value $\gamma_0=35 \text{ ns}^{-1}$ measured for Sample~1 in Ref.~\onlinecite{Shornikova2018ns}.}
	\label{fig:tauLfit}		
\end{figure}

In bare core CdSe NPLs, as well as in core/shell CdSe/ZnS NPLs the emission at low temperatures originates from excitons, see Fig.~\ref{fig:Fig1ab}a. Usually, the following three-level system is considered for describing the exciton recombination dynamics in colloidal NCs (Fig.~\ref{fig:tauLfit}a): the two lowest states of the exciton fine structure  with momentum projections $\pm 2$ (``dark'' excitons, dipole-forbidden transition $\ket{F}$) and $\pm 1$ (``bright'' excitons, dipole-allowed transition $\ket{A}$), and the unexcited crystal state $\ket{G}$. The bright and dark states are separated by the exchange energy $\Delta E_{\rm AF}$ of several meV. Within this model, the time-resolved PL at low temperatures is expected to show a bi-exponential decay with a short and a long-lived component with corresponding times $\tau_{\rm short}$ and $\tau_{\rm L}$. The short component is determined by radiative recombination of the bright exciton described by the rate $\Gamma_{\rm A}$ and its energy relaxation into the dark exciton state described by rate $\gamma_{0}$. The long component corresponds to the dark exciton recombination with rate $\Gamma_{\rm F}$ at low temperatures, when $kT \ll \Delta E_{\rm AF}$. At elevated temperatures, when the dark exciton can be activated to the bright state, $\Gamma_{\rm L}$ is a function of $\Gamma_{\rm A}$, $\Gamma_{\rm F}$ and $\Delta E_{\rm AF}$, which allows one to evaluate $\Delta E_{\rm AF}$ from the temperature dependence of $\Gamma_{\rm L}(T)$. This approach was discussed in detail in ref.~\onlinecite{Shornikova2018ns}, where Sample~1 and two other NPL samples are comprehensively analyzed.
The values of $\Gamma_{\rm L}$ evaluated from two-exponential fits to the data are shown by the filled red circles in Figs.~\ref{fig:tauLfit}b and \ref{fig:tauLfit}d. Their fits with equation~\eqref{eq:tauLongFull} are given by the dashed lines. 
\begin{equation}
\Gamma_{\rm L}(T) ={} \frac{1}{2} \left[ \Gamma_{\rm A} +\Gamma_{\rm F}+\gamma_0 \coth\left( \frac{\Delta E_{\rm AF}}{2kT} \right) - \sqrt{{\left( \Gamma_{\rm A} -\Gamma_{\rm F}+\gamma_0 \right)}^2+\gamma_0^2 \sinh^{-2}\left(\frac{\Delta E_{\rm AF}}{2kT} \right)} \right] .
\label{eq:tauLongFull}
\end{equation}

Most colloidal nanostructures can be treated by this approach. Even if small deviations from a bi-exponential behavior are present, they do not significantly affect the accuracy of the $\Delta E_{\rm AF}$ evaluation. Below we show that although Samples~1 and ~2 exhibit three-exponential decays, $\Delta E_{\rm AF}$ obtained from these fits are almost the same as from bi-exponential fits. Even Sample~3, which exhibits a multi-exponential behavior, can be also treated with this simple model.

\subsection{Temperature-dependent time-resolved PL in Samples~1 and~2} \label{subsec:SI_FStr_TR(T)_S1_S2}

In ref.~\onlinecite{Shornikova2018ns}, the PL decay of CdSe NPLs was fitted with a two-exponential function, which gives a reliable evaluation of the long decay time $\tau_{\rm L}$. In fact, the PL decay $I(t)$ across the whole temporal range is better reproduced by a three-exponential function:
\begin{equation}
I(t)=A_{\rm short}\exp(-t/\tau_{\rm short})+A_{\rm M}\exp(-t/\tau_{\rm M})+A_{\rm L}\exp(-t/\tau_{\rm L}) ,
\label{eq:3-exponential}
\end{equation}
where $\tau_{\rm short}$, $\tau_{\rm M}$, and $\tau_{\rm L}$ are the decay times, and $A_{\rm short}$, $A_{\rm M}$, and $A_{\rm L}$ are the amplitudes. Here $\tau_{\rm M}$ and $A_{\rm M}$ are the parameters describing the middle decay component.  

The real $\tau_{\rm short}$ value cannot be measured in our experiment that is limited by the time-resolution of the setup of 0.3~ns. A bright exciton lifetime of  $\tau_F= 22$~ps was measured for Sample~1 by a streak camera.\cite{Shornikova2018ns}
Here, to fit the measured PL decays, we take $\tau_{\rm short}=0.3$~ns, which is the response time of the avalanche photodiode to a 50~ps long laser pulse. 

The two other decay components arise from recombination of the dark excitons. As discussed in the main part, we assume that in the studied drop-casted samples there are two NPL subensembles: one with pure radiative recombination having the decay rate $\Gamma_{\rm L}=\tau_{\rm L}^{-1}$, and one with both radiative and nonradiative dark exciton recombination with decay rate $\Gamma_{\rm M}=\Gamma_{\rm L}+\Gamma_{\rm nr}=\tau_{\rm M}^{-1}$. Their integral intensities in zero field at $T=4.2$~K differ by about one order of magnitude, compare the values of 
$I_{\rm L} / (I_{\rm M} + I_{\rm L})$ (diamonds) and $I_{\rm M} / (I_{\rm M} + I_{\rm L})$ (circles) in Fig.~\ref{fig:SI_TR(B)}c.  This ratio does not vary with temperature (not shown). We will show in section~\ref{subsec:SI_TR(B)_Theor} that in an external magnetic field the increase of $\Gamma_{\rm L}$ reduces the relative contribution of $\Gamma_{\rm nr}$, as a result the integral intensities of $I_{\rm L}$ and  $I_{\rm M}$ approach each other. 

The middle component might be associated with NPLs in stacks. Indeed, it was reported that formation of stacked NPL ensembles results in a 10-fold decrease of the PL quantum efficiency compared to ensembles of isolated NPLs.\cite{Guzelturk2014} A possible reason is the energy transfer between NPLs in stacks, delivering excitons to nonradiative centers, \cite{Guzelturk2014} as the inter-platelet distance is only 5~nm,\cite{Tessier2013acs} and transfer rates of $6-10$~ps were reported.\cite{Rowland2015} Another possible nonradiative recombination channel for excitons can be provided by deep traps, surface imperfections, \textit{etc}. The existence of trap states was proven recently for CdSe NPLs.\cite{Tessier2012,Kunneman2014,Guzelturk2014,Olutas2015} 

The $\Gamma_{\rm M}$ and $\Gamma_{\rm L}$ from three-exponential fits are shown in Figs.~\ref{fig:tauLfit}b and \ref{fig:tauLfit}d by blue and green circles. One can see that taking into account the middle decay component allows us to obtain more precise decay rates $\Gamma_{\rm L}$ for temperatures below 7~K (compare the red and green circles). For higher temperatures, both fitting approaches give similar values of $\Gamma_{\rm L}$. The solid green lines are fits with equation~\eqref{eq:tauLongFull} of $\Gamma_{\rm L}(T)$. Since $\gamma_0$ is not known for Sample~2, we used for the fit the value $\gamma_0=35 \text{ ns}^{-1}$, measured for Sample~1 in Ref.~\onlinecite{Shornikova2018ns}. This is a reasonable assumption, since the order of magnitude of $\gamma_0$ does not change from sample to sample,\cite{Shornikova2018ns} and a small variation of $\gamma_0$ does not affect the fit. With the same parameters, we fit the middle component (blue lines) using $\Gamma_{\rm M}(T)=\Gamma_{\rm L}(T)+\Gamma_{\rm nr}$, where $\Gamma_{\rm nr}$ is the nonradiative decay rate, $\Gamma_{\rm nr}=\tau_{\rm nr}^{-1}=0.08$ ns$^{-1}$. For both Samples~1 and 2 the three-exponential fit gives a somewhat smaller value of $\Delta E_{\rm AF}=4.6\pm0.5$ meV compared to the two-exponential fit ($5.0\pm0.5$). We note that this value is closer to the $\Delta E_{\rm AF}=4.0\pm0.1$~meV obtained from fluorescence line narrowing measurements.\cite{Shornikova2018ns}   

\subsection{Temperature-dependent time-resolved PL in Sample~3} \label{subsec:SI_FStr_TR(T)_S3}

In the CdSe/ZnS NPLs (Sample 3) the CdSe core is sandwiched between ZnS layers with a large band gap of 3.7~eV. The emission linewidth of this sample is broader compared to the bare core CdSe NPLs (Fig.~\ref{fig:Fig1ab}a). The PL decay of the CdSe/ZnS NPLs is multi-exponential and at least four components are required for the fit. We attribute this behavior to contributions of both neutral excitons and negatively charged excitons (trions) to the PL. Our methods do not allow us to estimate the relative contributions of excitons and trions in the emission band. Nevertheless, exciton signatures are clearly seen in the time-resolved emission as an acceleration of the long-decay component at elevated temperatures (Fig.~\ref{fig:tauLfit}e,f) and in an external magnetic field (Fig.~\ref{fig:SI_TR(B)}b).

Fitting of the PL decays in Sample~3 is rather complicated, since it is hard to distinguish which of the decay components should be attributed to dark excitons. However, one can find an averaged long decay time, which characterizes the dynamics of about $80\%$ integral intensity for the long-lived components. The fits at temperatures of 2.2 and 70~K are shown in Fig.~\ref{fig:tauLfit}e by the dashed red lines.
The resulting $\Gamma_{\rm L}$ rates are shown in Fig.~\ref{fig:tauLfit}f. Two fits with equation~\eqref{eq:tauLongFull} with the same $\Gamma_{\rm A}=0.39$~ns$^{-1}$ and $\Delta E_{\rm AF}=3.9$~meV and two different $\Gamma_{\rm F}=0.003$ (green) and $0.002$~ns$^{-1}$ (blue) are shown. For the fitting, $\gamma_0=35 \text{ ns}^{-1}$ is used (see Section~\ref{subsec:SI_FStr_TR(T)_S1_S2}).
As one can see, below $T=6$~K a plateau at $\Gamma_{\rm L}=\Gamma_{\rm F}$ is expected in the modeling, while the experimental data show a further decrease with lowering temperature. 

Although the described fitting procedure is not accurate, it gives a reasonable $\Delta E_{\rm AF} = 3.9 \pm 0.5$~meV. This splitting energy is slightly smaller than that in bare CdSe NPLs with the same core thickness (Table I). This can be explained by the leakage of the electron wavefunction into the ZnS shell.\cite{Cruguel2017} Also, even if the exciton is well localized in the core, the dielectric confinement effect in core-shell structures is weaker, compared to the bare core NCs. Indeed, the 4-ML-thick CdSe cores are surrounded by 5-ML shells of ZnS, with about the same dielectric constant, \textit{i.e.}, the total NPL thickness increases more than threefold, and the dielectric screening is correspondingly increased. In the case of an infinitely thick shell, $\Delta E_{\rm AF}$ calculated within the effective mass approximation with account for the dielectric confinement falls in the range of $1.7-2.8$~meV, depending on the dielectric constants of the core and the shell (see Supplementary Figure~S8 in Ref.~\onlinecite{Shornikova2018ns}). Thus, $\Delta E_{\rm AF} = 3.9$~meV is a reasonable intermediate value between the splitting values for the bare core and infinite shell NPLs.

\section{Time-resolved emission in magnetic field} \label{sec:SI_TR(B)}

\subsection{Experimental data for Samples 2 and 3} \label{subsec:SI_TR(B)_Exper}

\begin{figure}[h]
	\includegraphics{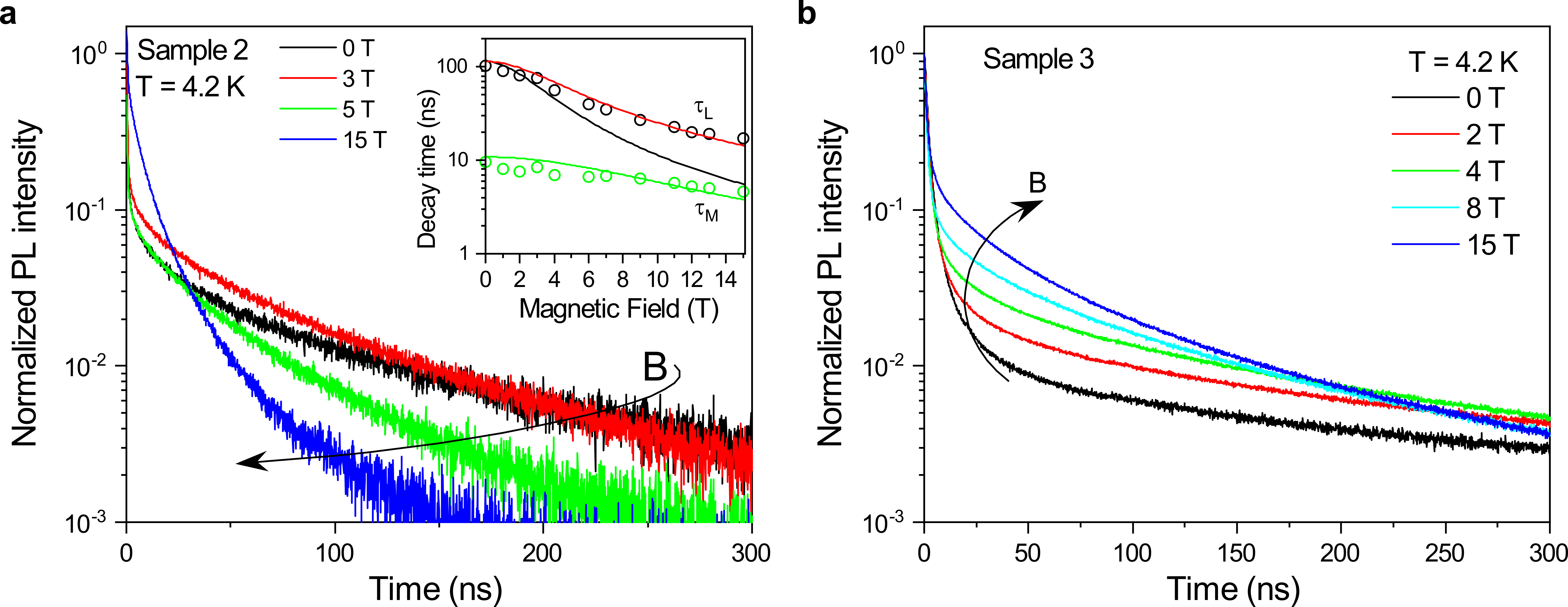}
	\caption{\textbf{PL decay of Samples~2 and~3 in various magnetic fields.} \textbf{a,} PL decays at maximum of exciton PL in Sample~2. Inset shows magnetic field dependences of times $\tau_{\rm M}$ (green circles) and  $\tau_{\rm L}$ (black circles). Green and black curves are decays calculated according to equation~\eqref{EQ:gammaFBM} with $\theta=\pi/2$, red curve is fit of $\tau_{\rm L}$ with $\theta=\pi/4$. \textbf{b,} PL decays at maximum of exciton PL in Sample~3. Acceleration of the long decay component $\tau_{\rm L}$ with increasing magnetic field is due to mixing of the dark and bright exciton states in NPLs whose normal is not parallel to the magnetic field direction.
		}
	\label{fig:TR(B)_Samples2and3}
\end{figure}

The exciton emission decay in colloidal NPLs is affected by the external magnetic field, which mixes bright and dark exciton states. Correspondingly, the long component shortens and its relative intensity increases, see Figs.~\ref{fig:Fig1ab}b, \ref{fig:TR(B)_Samples2and3}a and \ref{fig:TR(B)_Samples2and3}b. Interestingly, the dependence of the long component on magnetic field can be derived analytically (see below, Section~\ref{subsec:SI_TR(B)_Theor} and equation~\eqref{eq:Gver}).  The calculated dependence of the middle and long components of the PL decay, $\tau_{\rm M}$ and $\tau_{\rm L}$, for Sample~1 is shown in the inset of Fig.~\ref{fig:Fig1ab}b and for Sample~2 in the inset of Fig.~\ref{fig:TR(B)_Samples2and3}a.

The dependence of the time-resolved PL on temperature and magnetic field is also present in Sample~3, although it is less pronounced (Fig.~\ref{fig:TR(B)_Samples2and3}b). Apparently, we observe in this case PL signal both from excitons and trions (see discussion in Section~\ref{subsec:SI_FStr_TR(T)_S3}).
We associate that with partial photocharging of the CdSe/ZnS NPLs with electrons. This effect was well documented for CdSe/CdS spherical nanocrystals\cite{Liu2013} and for CdSe/CdS NPLs with thick shells exceeding 5~nm.\cite{Shornikova2018nl}     

\subsection{Modeling of recombination dynamics in magnetic field} \label{subsec:SI_TR(B)_Theor}

\begin{figure}[h!]
	\includegraphics[width=0.6\linewidth]{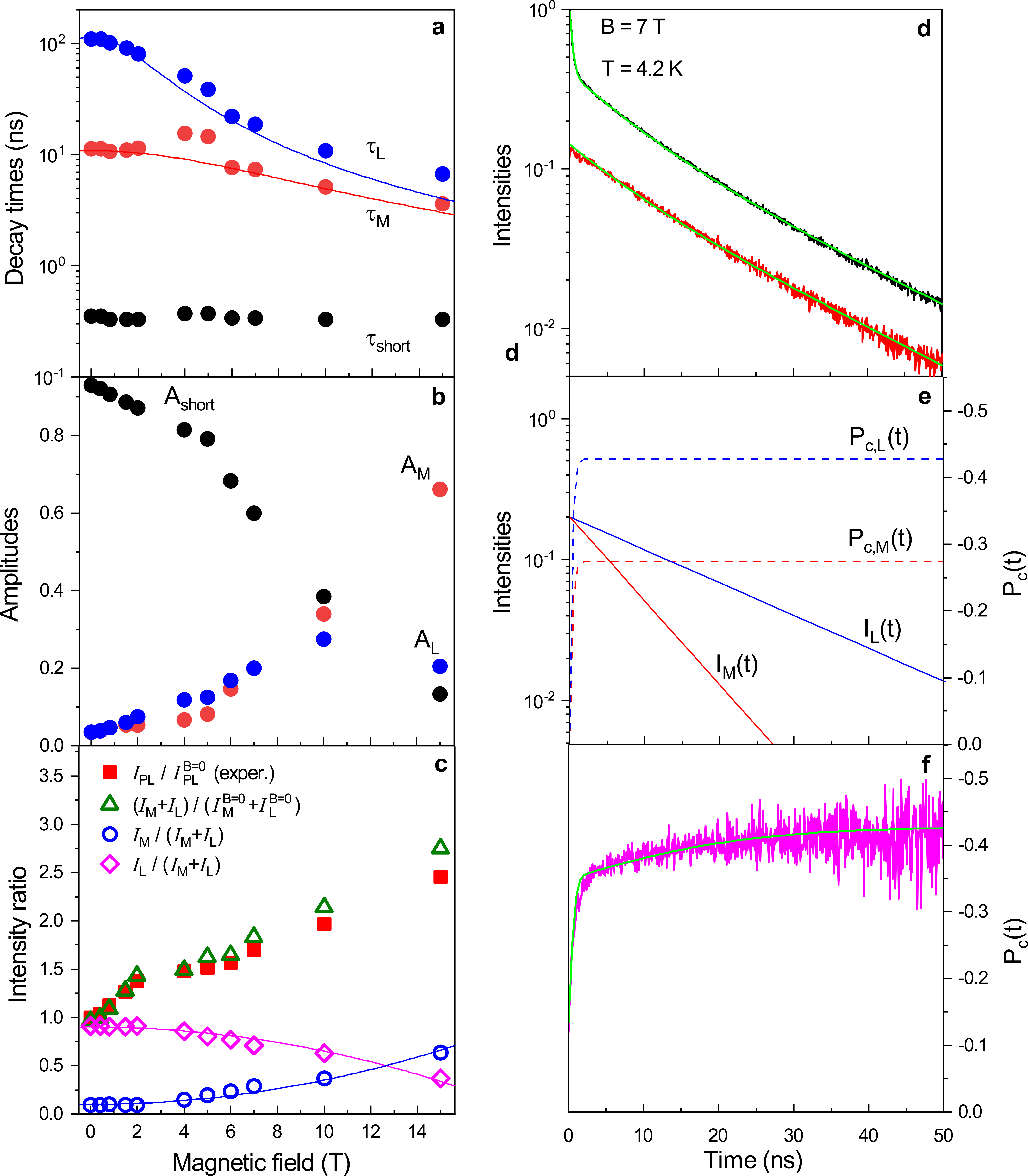}
	\caption{\textbf{Recombination dynamics of CdSe NPLs (Sample~1) in magnetic field at $\bf{\textit{T}=4.2}$~K.} Experimental data for times and amplitudes shown by symbols in the panels (a,b) are evaluated from three-exponential fits of recombination dynamics with  equation~\eqref{eq:3-exponential}. For parameters see text.
		\textbf{a}, Magnetic field dependencies of decay times. Lines are model calculations with equation~\eqref{EQ:gammaFBM}  for $\theta=\pi/2$. 
		\textbf{b}, Magnetic field dependencies of amplitudes. 
		\textbf{c}, Measured time-integrated PL intensity \textit{vs} magnetic field (red filled squares) normalized to its value in zero magnetic field. Open symbols show  calculated dependencies taking the times and amplitudes from panels (a, b) and using $I_{\rm M}=A_{\rm M}\tau_{\rm M}$ and $I_{\rm L}=A_{\rm L}\tau_{\rm L}$: normalized time-integrated intensity (green open triangles) and relative weights of $I_{\rm M}$ (open blue circles) and $I_{\rm L}$ (pink open diamonds). Blue line is fit according to $I_{\rm M}/(I_{\rm M}+I_{\rm L})=0.1+(B/20)^2$, and purple line is fit according to  $I_{\rm L}/(I_{\rm M}+I_{\rm L})=0.9-(B/20)^2$. These dependencies gives us the ratio $I_{\rm M}/I_{\rm L}=\left[0.1+(B/20)^2\right]/\left[0.9-(B/20)^2\right]$. 
		\textbf{d}, Experimental data and fits (green lines) for the dynamics of total PL intensity $I(t)=I^{+}(t)+I^{-}(t)$ (black) and the difference of polarized PL intensities $|I^{+}(t)-I^{-}(t)|$ (red). \textbf{e}, Intensities of middle and long decay components $I_{\rm M(L)}(t)=A_{\rm M(L)}\exp(-t/\tau_{\rm M(L)})$ (solid lines) and corresponding degrees of circular polarization $P_{\rm c,M(L)}(t)=\Delta A_{\rm M(L)}/A_{\rm M(L)}(1-\exp(-t/\tau_{\rm s}))$ (dashed lines). Here $\tau_{\rm s}$ takes into account spin relaxation which occurs over a subnanosecond period of time after the excitation pulse. 
		\textbf{f}, Comparison of temporal DCP dependencies  $[I^{+}(t)-I^{-}(t)]/[I^{+}(t)+I^{-}(t)]$ obtained from experiment (magenta) and three exponential fitting (green). }
	\label{fig:SI_TR(B)}
\end{figure}

Here we present and discuss the parameters of the recombination dynamics in Sample 1 measured in different magnetic fields up to 15~T at $T=4.2$~K. The decay of PL intensity in each magnetic field is fitted with the three-exponential function (equation~\eqref{eq:3-exponential}), introduced in Section~\ref{subsec:SI_FStr_TR(T)_S1_S2}. The obtained decay times and amplitudes are shown by the symbols in Figs.~\ref{fig:SI_TR(B)}a and \ref{fig:SI_TR(B)}b, respectively. One can see in Fig.~\ref{fig:SI_TR(B)}a that the long decay time $\tau_{\rm L}$ is shortened from 114~ns  down to 7~ns  with increasing magnetic field. This can be explained as shortening of the radiative recombination of the dark exciton due to admixture of the bright exciton. The blue line in Fig.~\ref{fig:SI_TR(B)}a shows a fit with equation~\eqref{EQ:gammaFBM} with $\theta=\pi/2$, corresponding to vertically aligned NPLs for which the magnetic field is applied parallel to the NPL plane in absence of nonradiative recombination ($\Gamma_{\rm nr}=0$). We use the following parameters, which validity has been explained before: $\Gamma_{\rm A}=10$~ns$^{-1}$,  $\Delta E_{\rm AF}=4.6$~meV, and $g_{\rm e}=1.7$. The good agreement with the experimental data for $\tau_{\rm L}(B)$ lets us conclude that the long decay component is provided by a subensemble of NPLs with vertical orientation that is not subject to nonradiative recombination. 

The magnetic field dependence of the middle decay time $\tau_{\rm M}=\Gamma_{\rm M}^{-1}$ is fitted also with equation~\eqref{EQ:gammaFBM}, but with account for a finite nonradiative recombination, that is taken into account through $\Gamma_{\rm M}(B)=\Gamma_{\rm L}(B)+\Gamma_{\rm nr}$, see Section~\ref{subsec:SI_FStr_TR(T)_S1_S2}. For the fit shown by the red line in Fig.~\ref{fig:SI_TR(B)}a we take $\Gamma_{\rm nr}=0.083$~ns$^{-1}$  and suggest that it is independent of the magnetic field strength. Again, the good agreement with the experimental data for $\tau_{\rm M}(B)$ allows us also here to conclude that the middle decay component is provided by a subensemble of NPLs with finite nonradiative recombination. 

Let us analyze the contributions of the different decay components to the time-integrated PL intensity $I_{\rm int}=\sum I_i=\sum A_i\tau_i$ in different magnetic fields. Here $A_i$ and $\tau_i$ ($i={\rm short}\text{, }{\rm M}\text{, }{\rm L}$) are the amplitudes and times given in Figs.~\ref{fig:SI_TR(B)}a and \ref{fig:SI_TR(B)}b. 
We find that the contribution of the short component to $I_{\rm int}$ is negligible across the whole range of magnetic fields.    
The intensity of the long component is nearly field-independent, it gives the major contribution to $I_{\rm int}$ in low magnetic fields, see the data for $I_{\rm L}/(I_{\rm M}+I_{\rm L})$ in Fig.~\ref{fig:SI_TR(B)}c. For the middle component, we observe a quadratic increase of the intensity, indicating an increase of the radiative quantum efficiency of the dark exciton in the subensemble with a finite nonradiative channel, due to mixing of the bright and dark exciton states. 
An important cross-check for the validity of our model description comes from the fact that the integral PL intensity determined from the recombination dynamics, $(I_{\rm M}+I_{\rm L})/(I_{\rm M}(B=0)+I_{\rm L}(B=0))$ shown by the open triangles in Fig.~\ref{fig:SI_TR(B)}c, coincides well with the time-integrated PL intensity, $I_{\rm PL}/I_{\rm PL}(B=0)$ shown by the filled squares. From that we conclude that the increase of the PL intensity in magnetic field is provided by the growing contribution of the M subensemble, for which the radiative recombination excels the nonradiative one as the field is increased.   

Let us turn now to the details of the  exciton spin dynamics in finite magnetic field. The results for Sample 1 at $B=7$~T are shown in Fig.~\ref{fig:SI_TR(B)}d. In the following we consider the DCP dynamics $P_{\rm c}(t)=[I^{+}(t)-I^{-}(t)]/[I^{+}(t)+I^{-}(t)]$. Experimentally the exciton recombination dynamics were measured separately for the $\sigma^+$ and $\sigma^-$ circular polarizations and the following forms calculated from them are given in Fig.~\ref{fig:SI_TR(B)}d: $I(t)=I^{+}(t)+I^{-}(t)$ (black), $|I^{+}(t)-I^{-}(t)|$ (red), and $P_{\rm c}(t)$ (magenta). Note that for the studied sample        
$I^{+}(t)<I^{-}(t)$ and $P_{\rm c}(t)<0$, but for convenience of presentation we plot $|I^{+}(t)-I^{-}(t)|$. Fits to the dynamics of these values are shown by the green lines. For $I(t)$ we use, as described above, equation~\eqref{eq:3-exponential} and get for $B=7$~T the following parameters, which are also shown in Figs.~\ref{fig:SI_TR(B)}a and \ref{fig:SI_TR(B)}b: $\tau_{\rm short}=0.33$~ns, $\tau_{\rm M}=7.2$~ns, $\tau_{\rm L}=18.2$~ns, $A_{\rm short}=0.6$, $A_{\rm M}=0.2$, and $A_{\rm L}=0.2$.
$|I^{+}(t)-I^{-}(t)|$ is fitted with a three-exponential function similar to equation~\eqref{eq:3-exponential}: 
\begin{gather}
|I^{+}(t)-I^{-}(t)|=\Delta A_{\rm short}\exp(-t/\tau _{\rm short})+\Delta A_{\rm M}\exp(-t/\tau _{\rm M}) +\Delta A_{\rm L}\exp(-t/\tau _{\rm L}).
\label{eq:dIdecay_SI}
\end{gather}
Here, the $\Delta A_i$ are the amplitudes of the decay components $|I^{+}(t)-I^{-}(t)|$ with corresponding decay times $\tau_i$. We use the times evaluated for $I(t)$ and get from the fit: $\Delta A_{\rm short}=0$, $\Delta A_{\rm M}=0.056$, and $\Delta A_{\rm L}=0.085$. 

This result shows that the component with lifetime $\tau_{\rm short}$ is absent in the decay of $|I^{+}(t)-I^{-}(t)|$, which is a consequence of the insufficient time resolution of the avalanche photodiode that does not allow to measure the intensities of the $\sigma^+$ and $\sigma^-$ polarized emission related the fast decay of the bright exciton. We also find that at the initial moment of time $t=0$ the difference $|I^{+}(0)-I^{-}(0)|$ for the middle and long decay components is non-zero, indicating fast spin relaxation between the dark exciton states $\ket{\pm2}$. For $t>0$, throughout the whole time range of PL decay of the middle $I_{\rm M}(t)$ and long $I_{\rm L}(t)$ components the corresponding values of DCP $P_{\rm c,M(L)}^{\rm eq}=P_{\rm c,M(L)}(t \to \infty)=\Delta A_{\rm M(L)}/A_{\rm M(L)}$ remain constant (see Fig.~\ref{fig:SI_TR(B)}d). The DCP of the long decay component is 1.5 times higher compared to the DCP of the middle decay component.
  
On the basis of these results we suggest the following interpretation of the DCP dynamics in Sample 1 (see Fig.~\ref{fig:SI_TR(B)}f). 
First, the increase of DCP in the temporal range from 0 to 1~ns is caused by the decay of the unpolarized emission of bright excitons. The further increase from 1 to 30~ns is specific for the situation when several subensembles (in our case two) with different equilibrium DCP contribute to the PL signal. 
Then the subensemble with the lower $P_{\rm c}^{\rm eq}$ will be depleted, resulting in an increase of $P_{\rm c}(t)$ during the recombination time of this subensemble. The good agreement of the model (green line) and the experimental data (magenta) for  $P_{\rm c}(t)$ shown in Fig.~\ref{fig:SI_TR(B)}f confirms the validity of the suggested explanation.
Note that one can also have the opposite situation, e.g., in case that the middle component has a larger DCP than the long one, the $P_{\rm c}(t)$ dynamics will show a decrease of the polarization degree with $\tau_{\rm M}$.

The experimental values of $P_{\rm c}^{\rm int}$ and $P_{\rm c}^{\rm eq}$ measured for Sample~1 at the  exciton maximum of $2.497$~eV (dashed line in Fig.~\ref{fig:Fig2abcd}b) are plotted in Fig.~\ref{fig:Fig2abcd}d.
For all studied temperatures and magnetic fields, the absolute values of $P_{\rm c}^{\rm int}$ are smaller than those of $P_{\rm c}^{\rm eq}$. Commonly one would expect that this finding is due to a difference between the exciton lifetime $\tau$ and the spin relaxation time $\tau_{\rm s}$: if $\tau_{\rm s}$ is long enough, the dynamical factor $\tau/(\tau+\tau_{\rm s})<1$, do that $|P_{\rm c}^{\rm int}|=  |P_{\rm c}^{\rm eq}|\tau/(\tau+\tau_{\rm s})< |P_{\rm c}^{\rm eq}|$.\cite{Liu2013}  However, as we have shown above, this is not the case for the studied samples.   This allows us to conclude that the slow increase of $|P_{\rm c}(t)|$ with time (Fig.~\ref{fig:SI_TR(B)}f), as well as the relation $|P_{\rm c}^{\rm int}|<|P_{\rm c}^{\rm eq}|$, comes from different DCP in the two NPL subensembles:  $P_{\rm c,M}<P_{\rm c,L}$. This also means that the equilibrium intensities $I^\pm$ in equation (\ref{eq:DCP_definition}) correspond to the intensities at $t > \tau_{\rm M}$, in agreement with Fig.~\ref{fig:SI_TR(B)}d,e,f.

\section{Supplementary data on DCP in magnetic field \label{sec:SI_DCP(B)}} \label{sec:SI_Suppl_DCP}

\textbf{Hole $g$ factor.} The hole $g$ factor can be determined from the DCP dependence on magnetic field for the negatively charged trion (low-energy peak) shown in Fig.~\ref{fig:Pint(B)_3samples}b. We showed in ref.~\onlinecite{Shornikova2018nl} that in CdSe NPLs the type of charged exciton state, i.e., whether it is negatively or positively charged, can be unambiguously determined from the DCP sign. Negative DCP, as we have in our case, corresponds to emission from negatively charged excitons. In this case the two contributing electrons are in a singlet state with opposite spin orientations and therefore do not contribute to the trion DCP, which is determined by the polarization of the hole in the trion. The exact values of the hole $g$ factor $g_h$, which governs the Zeeman splitting of the trion state (but not the splitting of the corresponding optical transition!), is hard to evaluate from the data of Fig.~\ref{fig:Pint(B)_3samples}b, as the $P_{\rm c}(B)$ dependence there does not reach saturation. Supposing that $P_{\rm c}^{\rm sat}=-0.75$, we obtain for both Samples 1and 2 $g_h=-0.03$. In case of $P_{\rm c}^{\rm sat}=-0.2$ one gets $g_h=-0.1$. These estimations are made using equation (2), where the hole Zeeman splitting of $3g_h\mu_{\rm B}B$  was used instead of $\Delta E_{\rm F}$.  The latter value is in accordance with the theoretically calculated $g_h$ for a cubic zinc-blende CdSe quantum well with parabolic potential.\cite{Shornikova2018nl}

\begin{figure}[h!]
	\includegraphics{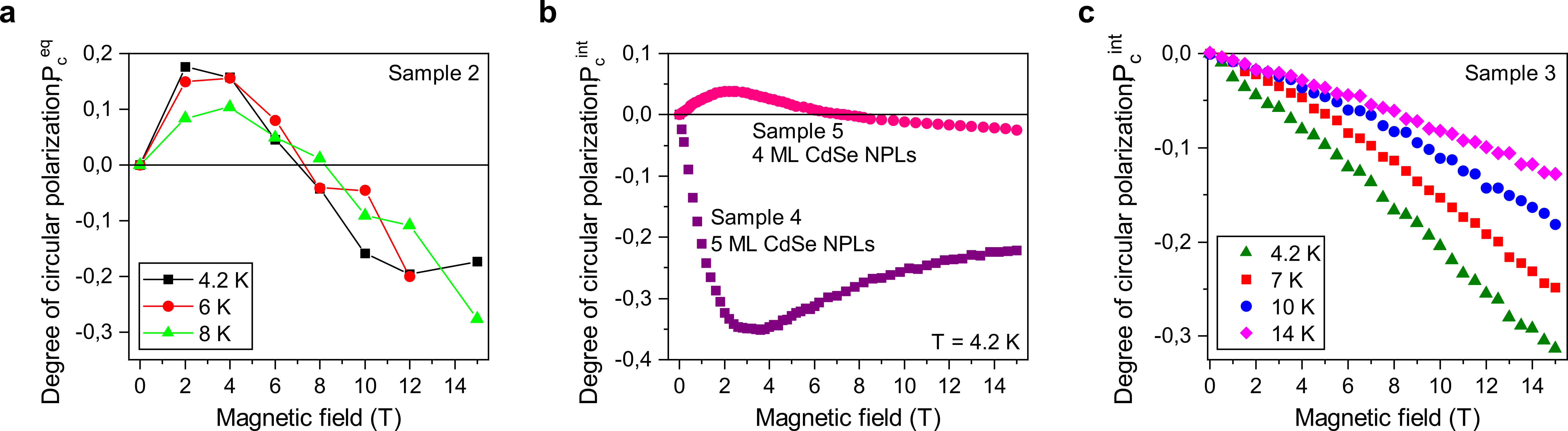}
	\caption{\textbf{Additional data on DCP in magnetic field.}
\textbf{a}, Magnetic field dependence of the time-saturated DCP, $P_{\rm c}^{\rm eq}$, in Sample~2 at various temperatures. Similar to the $P_{\rm c}^{\rm int}$ dependence in Fig.~\ref{fig:Pint(B)_3samples}c, $B_c$ increases with growing temperature. \textbf{b}, Non-monotonous DCP dependence on magnetic field measured for two more NPL samples, which are not included in the main text: (i) Sample~4: 5-monolayers CdSe NPLs (described in detail in Ref.~\onlinecite{Shornikova2018ns}), grown in argon atmosphere; (ii) Sample 5: 4-monolayers CdSe NPLs, similar to Sample~2 but with different lateral dimensions: $9\times 9$~nm$^2$, grown in ambient conditions. \textbf{c}, Magnetic field dependence of the time-integrated DCP, $P_{\rm c}^{\rm int}$, at various temperatures for CdSe/ZnS NPLs (Sample~3) measured at the PL maximum.}
	\label{fig:SI_Fig5}
\end{figure}

\textbf{$P_{\rm c}^{\rm eq}$ at various temperatures in Sample~2.}
We suggested that the $P_{\rm c}^{\rm eq}(B)$ measured at different temperatures gives insight into the mechanism responsible for the radiative recombination of the dark excitons, see the modeling in Figs.~\ref{fig:Fig4_theory_main}c and \ref{fig:Fig4_theory_main}d. Namely, the temperature dependence of the critical magnetic field $B_{\rm c}$, at which $P_{\rm c}^{\rm eq}(B_{\rm c})=0$, can be used to distinguish between spin-independent and spin-dependent recombination of the dark excitons. 
In the first case, $B_{\rm c}$ corresponds to zero Zeeman splitting of the dark exciton $\Delta E_{\rm F}=0$, \textit{i.e.} at this field the intrinsic Zeeman splitting is fully compensates by the splitting induced by the exciton exchange interaction with surface spins. A temperature increase reduces the surface spin polarization in finite magnetic field and, therefore, reduces the exchange interaction, while the intrinsic Zeeman splitting remains the same. As a result, $B_{\rm c}$ should decrease with increasing temperature. This case in modeled in  Fig.~\ref{fig:Fig4_theory_main}c. 
In the second case of strong spin-dependent recombination of the dark excitons and of an antiferromagnetic type of exchange interaction, which is realized in Sample~2, $B_{\rm c}$ should shift to larger values with increasing temperature, see Fig.~\ref{fig:Fig4_theory_main}d.  
In Sample 2 $B_{\rm c}$ is shifted to larger vales, as one can see both for $P_{\rm c}^{\rm int}$ (Fig.~\ref{fig:Pint(B)_3samples}c) and $P_{\rm c}^{\rm eq}$ (Fig.~\ref{fig:SI_Fig5}a). Therefore, we conclude that the spin-dependent recombination of the dark excitons, assisted by their exchange interaction with surface spins, plays an important role. Note that the data in Fig.~\ref{fig:SI_Fig5}a were measured on another spot of the dropcasted sample, so that the fraction of horizontal and vertical NPLs is different than in the data presented in Fig.~\ref{fig:Pint(B)_3samples}c. Therefore, one should not compare these two data sets directly. 

\textbf{Non-monotonous DCP dependence on magnetic field in CdSe NPLs (Samples 4 and 5).}
In order to demonstrate that the experimental appearances of the NPL polarization properties are not individual features of specific samples, but can bereproduced for various synthesis methods, we give in  Fig.~\ref{fig:SI_Fig5}b two more examples of non-monotonous $P_{\rm c}^{\rm int}(B)$ for Samples 4 and 5. 

Sample 4 contains bare core CdSe NPLs with a thickness of 5 monolayers, synthesized in argon atmosphere similar to Sample~1, their surface properties should be also similar. This sample was characterized by optical methods in detail in ref.~\onlinecite{Shornikova2018ns}. Its emission is dominantly $\sigma^{-}$-polarized, reaching a polarization degree of $-0.35$ at $B=3$~T which is very similar to Sample 1 with 4-monolayer-thick NPLs, see the red squares in  Fig.~\ref{fig:Pint(B)_3samples}a.

Sample~5 contains bare core 4ML NPLs, synthesized in ambient conditions following a protocol similar to Sample~2, but with different lateral dimensions: $9\times 9$~nm$^2$. 
As may be expected, sample~5 exhibits a behavior similar to Sample~2: the DCP is positive in small magnetic fields, crosses zero at $B=7.4$~T and decreases to $-2.5\%$ at $B=15$~T, compare with the blue circles in Fig.~\ref{fig:Pint(B)_3samples}a. 

\textbf{Magnetic field dependence of the time-integrated DCP in CdSe/ZnS NPLs (Sample~3).}
Figure~\ref{fig:SI_Fig5}c shows the magnetic field dependence of the time-integrated DCP, $P_{\rm c}^{\rm int}$, at various temperatures for the CdSe/ZnS NPLs (Sample~3) measured at the PL maximum. The green triangles repeat the data from Fig.~\ref{fig:Pint(B)_3samples}a. Unlike for bare core CdSe NPLs, the DCP of the CdSe/ZnS NPLs is a monotonous function of magnetic field and shows an about linear dependence over the whole field range up to 15~T for temperatures between 4.2 and 14~K. The decrease of DCP with increasing temperature is typical for a thermal population of the Zeeman sublevels. This behaviour and the much smaller slope of the  $P_{\rm c}^{\rm int}$ dependence evidence that surface spins in CdSe/ZnS NPLs are either absent or sufficiently separated from the excitons localized in CdSe cores by the ZnS shells. Therefore, their exchange contribution to the Zeeman splitting of the dark excitons is negligible.

\section{Theory \label{sec:theory}} \label{sec:SI_Theory}

\subsection{Degree of circular polarization in bi-modal ensemble of nanopletelets} \label{subsec:SI_Theor_DCP_bi-modal}

The circular polarization of the PL in external magnetic field arises from different populations of the dark exciton Zeeman sublevels $\ket{\pm2}$. The difference in occupation probabilities $p_{\rm F}^{\pm}$ of these sublevels is determined by the exciton Zeeman splitting $\Delta E_{\rm F}=E_{+2}-E_{-2}$. This Zeeman splitting $\Delta E_{\rm F}$ depends on the exciton $g$ factor, $g_{\rm F}=g_e-3g_h$, which is determined by the $g$ factors of the electron, $g_{\rm e}$, and the hole, $g_{\rm h}$. In NPLs, the total angular momentum of the hole is aligned along the quantization axis due to the strong shape anisotropy. In low magnetic fields, the electron spin also remains pinned along the quantization axis due to the strong exchange interaction between electron and hole in the exciton. As a result, the exciton Zeeman splitting depends on the angle $\theta$ between the quantization axis and the magnetic field direction, because $\Delta E_{\rm F}=g_{\rm F}\mu_{B}B \cos \theta$.\cite{JohnstonHalperin2001,Shornikova2018nl} 

In our experiments, the magnetic field was applied perpendicular to the plane of the substrate on which NPLs were deposited. Due to the parallelepiped shape of the NPLs, they tend to have preferable orientations when drop-cast on a substrate. First, single NPLs are arranged mostly horizontally, as one may expect (see Fig.~\ref{fig:TEM}a). The anisotropy axis in these NPLs is directed parallel to the magnetic field that is applied in the Faraday geometry, i.e. $\theta = 0$, so that maximum Zeeman splitting is achieved. 
Stacked NPLs are usually arranged vertically on the substrate, i.e., their anisotropy axis is oriented perpendicular to the magnetic field with $\theta=\pi/2$. Due to that, the dark exciton states $\ket{\pm2}$ remain degenerate in magnetic field for these vertical NPLs.   

For the sake of clarity, in the following we neglect any small inclination of the NPLs from the horizontal and vertical orientation and consider a bi-modal distribution function $f_{\rm or}(\theta)$ of the NPL orientations in the ensemble: $f_{\rm or}(\theta)= n_{\rm hor} \delta(1-\cos \theta) + n_{\rm ver} \delta(\cos \theta)$. Here $n_{\rm hor}$ and  $n_{\rm ver}$ are the fractions of the horizontally and vertically oriented NPLs, respectively, that can be excited by laser, with $n_{\rm hor} + n_{\rm ver} = 1$. 
For such a bi-modal ensemble, the difference of polarized PL intensities $|I^{+}-I^{-}|$ comes from the horizontally oriented NPLs, while the total PL intensity  $I=I_{\rm }^{+}+I^{-}$ is contributed by both the horizontal and vertical NPLs. The resulting DCP is given by
\begin{equation}
P_{\rm c}(B)=\frac{I_{\rm hor}(B) }{I_{\rm hor}(B)+I_{\rm ver}(B)} P_{\rm c}^{\rm hor}(B)\, ,
\label{eq:PcB_SI}
\end{equation}
where $I_{\rm hor}=I_{\rm  hor}^{+}+I_{\rm hor}^{-}$ and $I_{\rm ver}=I_{\rm ver}^{+}+I_{\rm ver}^{-}$ are the total intensities of the PL from the horizontally and vertically oriented NPLs, respectively, and $P_{\rm c}^{\rm hor}(B)=(I_{\rm hor}^{+}-I_{\rm  hor}^{-})/I_{\rm hor}$. Equation~(\ref{eq:PcB_SI}) is also valid for an ensemble of vertical and horizontal NPLs with negative trion emission as observed for thick-shell CdSe/CdS NPLs.\cite{Shornikova2018nl} Indeed, the Zeeman splitting of the negative trion is determined by the splitting of the heavy hole spin sublevels $\Delta E_{\rm hh}=-3g_{\rm h}\mu_BB\cos\theta$. With this knowledge one can determine the ratio of vertical and horizontal CdSe/CdS NPLs deposited on the substrate.

\subsection{DCP of horizontally oriented NPLs} \label{subsec:SI_Theor_DCP_hor} 

The very steep increase of the DCP in Sample 1 and the positive DCP in Sample 2 in low  magnetic fields evidence that the Zeeman splitting of the dark excitons has a further contribution in addition to the intrinsic one. This contribution depends on the NPL synthesis conditions and thus it is logical to relate it to the NPL surface, namely to the presence of surface spins which are interacting with the spin of the dark exciton.  We suggest that the spin polarization of the unpassivated dangling bonds at the NPL surface can be the origin of the exchange field, which contributes to the exciton Zeeman splitting in the following way:
\begin{eqnarray}
\Delta  E_{\rm F}=g_{\rm F}\mu_{B}B +  2 E_{\rm p}  \rho_{\rm db}  \, ,
\label{eq:Zdb_SI}
\end{eqnarray}
where $\rho_{\rm db} = (n_{\rm db}^{-}-n_{\rm db}^{+})/n_{\rm db}=\tanh(g_{\rm db}\mu_BB/2kT)$ is the polarization of the dangling bond spins (DBSs), $n_{\rm db}^{+}$ ($n_{\rm db}^{-}$) is the surface density of the dangling bonds with spin parallel (antiparallel) to the magnetic field direction, $n_{\rm db}=n_{\rm db}^{+}+n_{\rm db}^{-}$ being the surface density of dangling bonds, $E_{\rm p}$ is the energy of exchange interaction of exciton with dangling bonds,  and $g_{\rm db}$ is the dangling bond $g$ factor.  In colloidal NCs electron has typicaly much larger density of its wavefunction at the surface, compared to hole. In this case the exciton-dangling bond interaction is provided by the electron-dangling bond interaction only being $E_{\rm p}=\alpha N_{\rm db}$, where $\alpha$  is the average electron-DBS exchange constant and $N_{\rm db}$ is the number of dangling bonds interacting with the electron composing exciton.~\cite{Rodina2018JEM} 
In nanoplateletes, $N_{\rm db}=2n_{\rm db} S_{\rm loc}$, with $S_{\rm loc}$ being the area of the electron (exciton) lateral localization. The average exchange constant $\alpha$ is defined as $\alpha=a_{0}^3\tilde \alpha|\psi_e(\pm L/2)|^2/S_{\rm loc}= \alpha_{\rm 2D}/S_{\rm loc}$~\cite{Rodina2018JEM}, where $a_{0}^3$ is the unit cell volume, $\tilde \alpha$ is the short-range exchange constant, and $|\psi_e(\pm L/2)|^2$ is the square of the electron wavefunction at the NPL surface. 

We assume that $\rho_{\rm db}$ is described by the equilibrium polarization of surface spins with $g$ factor $g_{\rm db}=2$, which is typical for surface paramagnetic centers in CdSe \cite{Ditina1968}: $\rho_{\rm db}=\tanh ({g_{\rm db}\mu_{B}B}/{2kT})$.
In this case the exciton Zeeman splitting in low magnetic fields ($g_{\rm db}\mu_{B}B < 2kT$) is given by 

\begin{eqnarray}
\Delta  E_{\rm F}(B) \approx (g_{\rm F}+E_{\rm p}g_{\rm db}/kT) \mu_{\rm B}B \, .
\label{eq:Zdblin_SI}
\end{eqnarray}

If $g_{\rm F}$ and $E_{\rm p}g_{\rm db}$ have the same signs, the exchange field of surface spins increases the exciton Zeeman splitting. In case of opposite signs, it decreases the Zeeman splitting and can even change its sign if the condition $|g_{\rm F}|<|E_{\rm p}g_{\rm db}/kT|$ is fulfilled. 

 The dependencies of $\Delta E_{\rm F}(B)$ calculated according to equation \eqref{eq:Zdb_SI} on magnetic field  at $T=4.2$~K and on temperature at $B=4$~T are shown in Supplementary Fig.~\ref{fig:dEf_SI}a,b for different values of $E_{\rm p}$. If the magnitude of negative $E_{\rm p}$ is large enough $\Delta E_{\rm F}(B)$ changes its sign in low magnetic fields. Increase of  the magnetic field at a given temperature or increase of the temperature at a given magnetic field turns the splitting $\Delta E_{\rm F}(B)$  to zero at $B_0=2|E_{\rm p}|\rho_{\rm db}/g_{\rm F}\mu_{\rm B}$ (Fig.~\ref{fig:dEf_SI}a,b). Increase of the temperature or decrease of the $E_{\rm p}$ magnitude shift the compensating magnetic field $B_0$ to lower values (Fig.~\ref{fig:dEf_SI}c,d ).For modelling we take exciton $g$ factor $g_{\rm F}=g_{\rm e}-3g_{\rm h}\approx2$, being composed of the electron and hole $g$ factors: $g_{\rm e}\approx1.7$\cite{Kalitukha2018} and $g_{\rm h}\approx-0.1$. This hole $g$-factor value is in agreement with the rise of the trion DCP in Sample 1 (Supplementary Section~\ref{sec:SI_Suppl_DCP}), as well as with the theoretically calculated value \cite{Shornikova2018nl}.

\begin{figure}[h!]
	\includegraphics[width=0.8\linewidth]{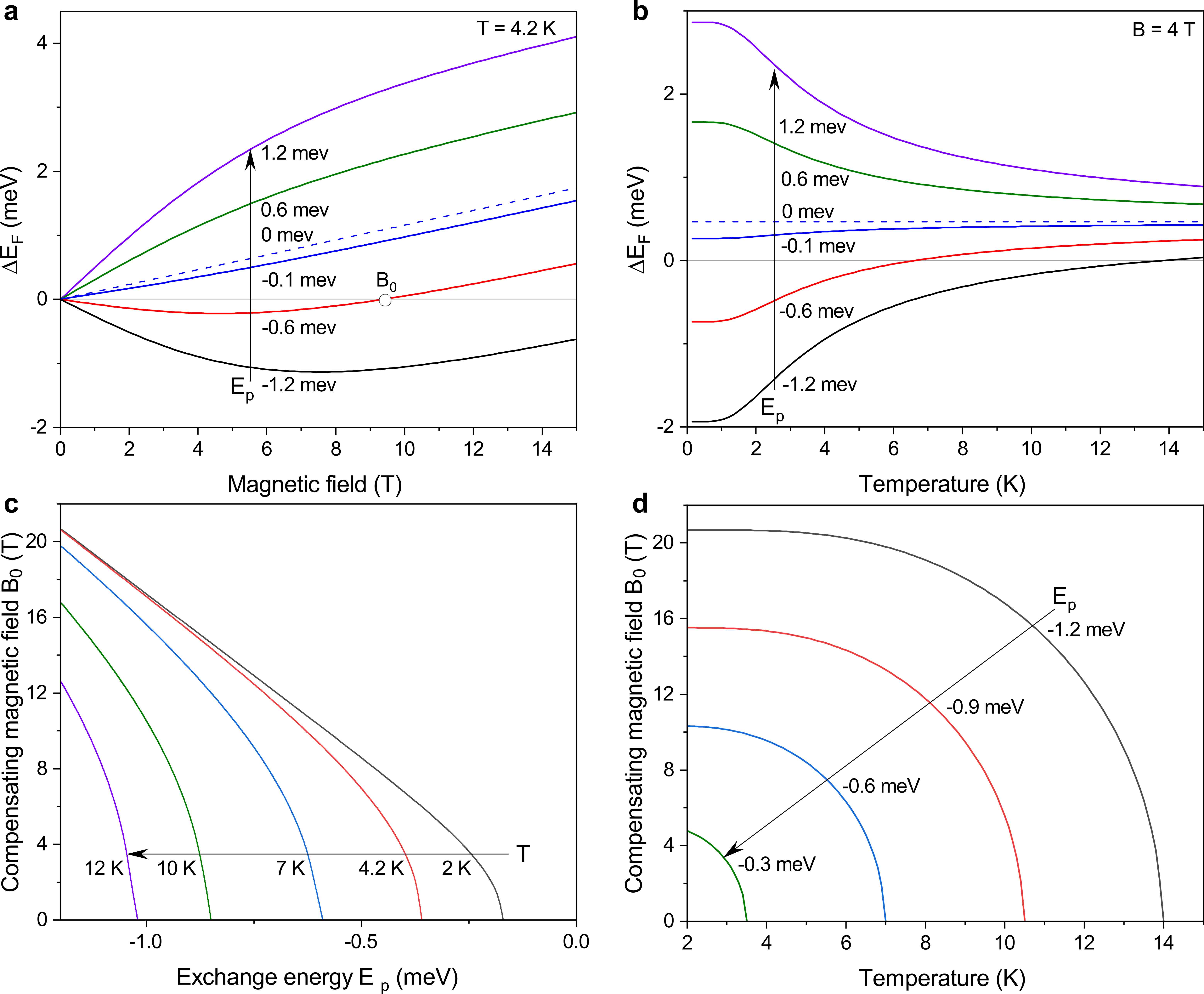}
	\caption{Exciton Zeeman splitting $\Delta  E_{\rm F}=g_{\rm F}\mu_{B}B +  2 E_{\rm p}  \rho_{\rm db}$ calculated as  a function of  magnetic field at $T=4.2$~K (panel \textbf{a}) and as a function of temperature at $B=4$~T  (panel \textbf{b}) for different values of $E_{\rm p}=-1.2$, $-0.6$, $-0.1$, $0$, $0.6$, and $1.2$~meV. Dependence of the compensating magnetic field $B_0=2|E_{\rm p}|\rho_{\rm db}/g_{\rm F}\mu_{\rm B}$ from $E_{\rm p}$ at different temperatures  (panel \textbf{c}) and from temperature at different$E_{\rm p}$ (panel \textbf{d}). In all panels we use $g_{\rm F}=g_{\rm db}=2$.}
	\label{fig:dEf_SI}
\end{figure}

Interaction with the surface spins can also influence the recombination dynamics of the dark exciton. Indeed, as the dark exciton radiative recombination is forbidden within the electric-dipole approximation, it requires activation mechanisms that provide an admixture of the bright exciton. One of the most efficient activation mechanisms is the flip of the electron spin via interaction with a dangling bond spin.\cite{Rodina2015,Rodina2016} The efficiency of this mechanism depends on the number of DBSs interacting with the electron and having spin orientation opposite to the electron spin. When the DBSs have no preferential orientation, e.g. are unpolarized at zero magnetic field, the DBS-assisted recombination rate of the  dark exciton is given by $\Gamma_{\rm db}= 2\gamma_{\rm ex}N_{\rm db}$, where $2\gamma_{\rm ex}$  describes the probability of the simultaneous flip-flop of the electron spin and one DBS. The explicit expression for $\gamma_{\rm ex}$ is given in Ref.~\onlinecite{Rodina2018JEM}. If the DBSs are polarized, for example in an external magnetic field, the radiative recombination rates of the $\ket{\pm2}$ exciton states become different. For horizontal NPLs these rates are equal to
\begin{eqnarray}
\Gamma_{\rm hor}^{\pm{2}}(B)=\Gamma_{\rm F}(B) +
\Gamma_{\rm db}[1\pm\rho_{\rm db}(B)] \, ,
\label{eq:Gpm2_SI}
\end{eqnarray}
where  $\Gamma_{\rm F}(B)$ is the dark exciton radiative recombination rate not assisted by the dangling bonds. This results in
\begin{eqnarray}
P_{\rm c}^{\rm hor}=\frac{p^{+}_\text{F}\Gamma_{\rm hor}^{+2}-p^{-}_\text{F}\Gamma_{\rm hor}^{-2}}
{p^{+}_\text{F}\Gamma_{\rm hor}^{+2}+p^{-}_\text{F}\Gamma_{\rm hor}^{-2}}=\frac{-\rho_{\rm ex}(\Gamma_{\rm F}+\Gamma_{\rm db})+\rho_{\rm db}\Gamma_{\rm db}}
{\Gamma_{\rm F}+\Gamma_{\rm db}(1-\rho_{\rm ex}\rho_{\rm db})}  \, .
\label{eq:Pchor_SI}
\end{eqnarray}
Here $\rho_{\rm ex}= (p_{\rm F}^{-}-p_{\rm F}^{+})/(p_{\rm F}^{-}+p_{\rm F}^{+})$ is the exciton polarization.
One can see that the degree of circular polarization of PL $ P_{\rm c}^{\rm hor}= -\rho_{\rm ex}$ only in the case $\rho_{\rm db}\Gamma_{\rm db}=0$.

In the general case the occupation probabilities $p_{\rm F}^{\pm}$ of the
dark exciton spin sublevels can be found from solving a system of coupled rate equations, taking into account the spin relaxation rates $\gamma_{\rm F}^\pm$ between the $\ket{\pm 2}$ exciton states:
\begin{eqnarray}
\frac{dp^{+}_\text{F}}{dt} &=& -{p^+_\text{F}}(\Gamma_{+2}+\gamma^+_\text{F})+ p_\text{F}^{-}\gamma_\text{F}^{-} + G(t)\, , \\
\frac{dp^{-}_\text{F}}{dt} &=& -{p^{-}_\text{F}}(\Gamma_{-2}+\gamma^{-}_\text{F})+ p_\text{F}^{+}\gamma_\text{F}^{+} +G(t) \, . \nonumber \label{eq:pop_SI}
\end{eqnarray}
Here  $\Gamma_{\pm2}$ are the total recombination rates of the dark excitons including nonradiative recombination, $G(t)$ is the dark exciton generation rate. 
As the spin relaxation is very fast in the samples under investigation, one can consider equilibrium populations of the spin sublevels $\ket{\pm2}$ determined by the Boltzmann statistics $\rho_{\rm ex}=\tanh(\Delta E_{\rm F}/2kT)$. 
In the case of a small exciton-DBS exchange splitting ($2E_{\rm p} \rho_{\rm db} \ll g_{\rm F} \mu_{\rm B} B$) and $g_{\rm F}<g_{\rm db}$ the spin-dependent recombination controls the DCP sign.

\begin{figure}[h!]
	\includegraphics[width=0.8\linewidth]{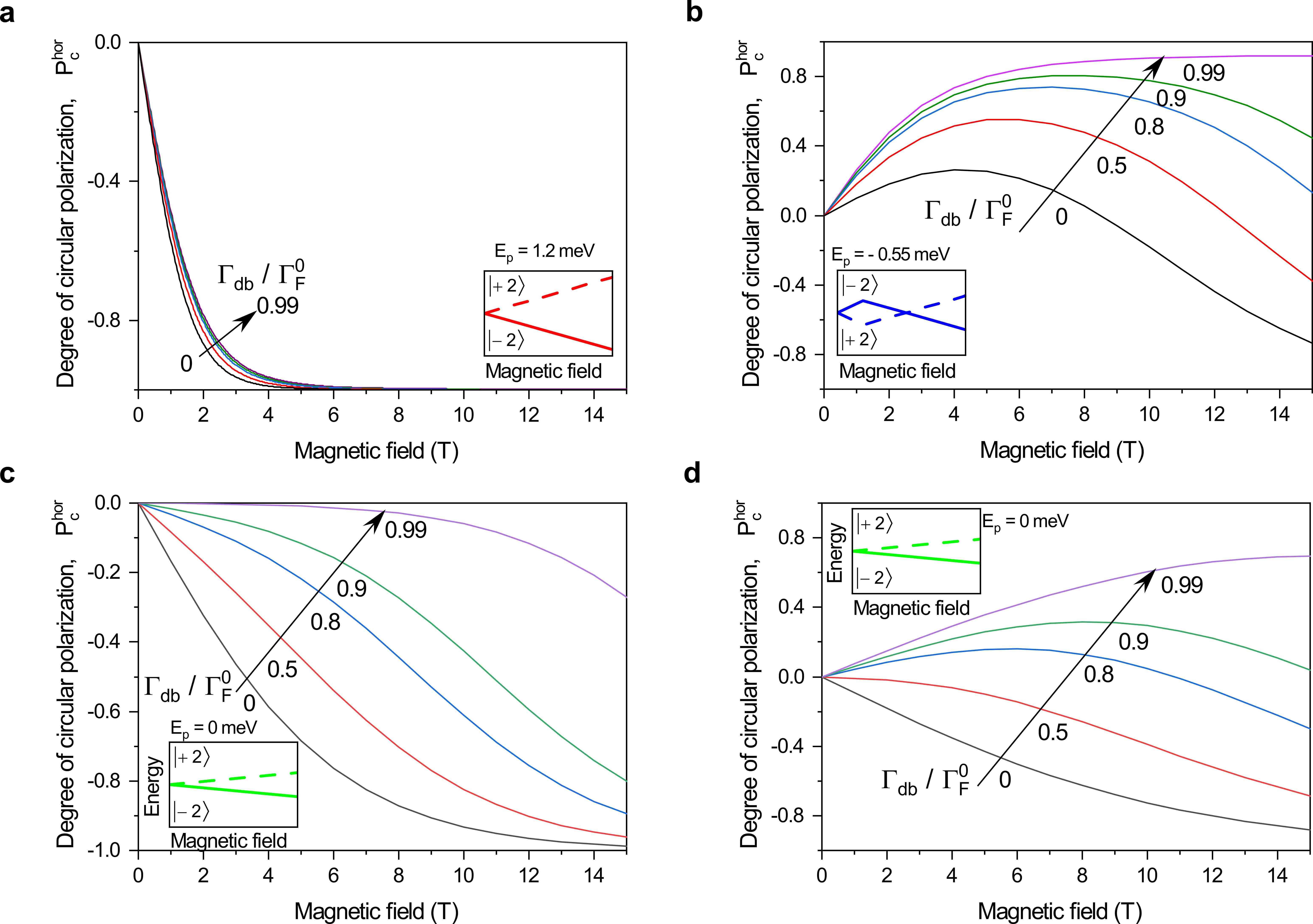}
	\caption{Dependence of equilibrium DCP in ensemble of horizontal NPLs with different $\Gamma_{\rm db}/\Gamma_{\rm F}^0$ ratio. \textbf{a}, $E_{\rm p}=1.2$ meV and  $g_{\rm F}=g_{\rm db}=2$. \textbf{b}, $E_{\rm p}=-0.55$ meV and  $g_{\rm F}=g_{\rm db}=2$. \textbf{c}, $E_{\rm p}=0$ meV and  $g_{\rm F}=g_{\rm db}=2$. \textbf{d}, $E_{\rm p}=0$ meV and  $g_{\rm F}=1.1<g_{\rm db}$ with $g_{\rm db}=2$. Inserts show schematically the respective exciton Zeeman splitting.}	
	\label{fig:SIEdbGdb}
\end{figure}

As shown in Fig.~\ref{fig:Fig4_theory_main}, a strong antiferromagnetic exchange interaction between the dark exciton and DBSs or a moderate antiferromagnetic exchange interaction together with spin-dependent recombination results in an inversion of the DCP sign. However, even in the case of negligibly small exciton-DBS exchange splitting ($2E_{\rm p} \rho_{\rm db} \ll g_{\rm F} \mu_{\rm B} B$) or compensation by simultaneous interaction of ferromagnetic and antiferromagnetic type the sign of the DCP can be inverted. This requires strong dangling bond assisted recombination of the dark exciton ($\Gamma_{\rm db} \gg \Gamma_{\rm F}$) and fulfillment of the condition $g_{\rm F}<g_{\rm db}$ (see Fig.~\ref{fig:SIEdbGdb}d). In this case the DBSs do not contribute to the Zeeman splitting of the dark exciton, which is $\Delta  E_{\rm F}=g_{\rm F}\mu_{B}B$, see equation~\eqref{eq:Zdb_SI} for $E_{\rm p}=0$. Also the dark excitons have a thermal equilibrium population of the spin sublevels due to the very fast spin relaxation, but the recombination probability from the exciton spin levels is different. From equation~\eqref{eq:Pchor_SI} we find that the condition for DCP sign inversion in low magnetic fields is $g_{\rm db}=g_{\rm F}(1+\Gamma_{\rm F}/\Gamma_{\rm db})$. If $g_{\rm F}\geq g_{\rm db}$ the sign of the DCP always remains the same due to the relation $\rho_{\rm ex}(\Gamma_{\rm F}+\Gamma_{\rm db})>\rho_{\rm db}\Gamma_{\rm db}$  (see Fig.~\ref{fig:SIEdbGdb}a).

\subsection{Intensities and radiative rates of dark excitons in horizontal and vertical nanoplatelets} 
\label{subsec:SI_Theor_Intensities_Rates}

According to equation~\eqref{eq:PcB_SI} the degree of circular polarization of PL from a bi-modal ensemble depends also on the ratio $I_{\rm hor}(B)/\left[I_{\rm hor}(B)+I_{\rm ver}(B)\right]$. The intensities $I_{\rm hor}(B,t)$ and $I_{\rm ver}(B,t)$ from the two subensembles with lifetimes $\tau_{\rm hor}$ and $\tau_{\rm ver}$  can be expressed as
\begin{equation}
\begin{aligned}
I_{\rm hor}(B,t) = 2n_{\rm hor}\Gamma_{\rm hor}(B)\exp(-t/\tau_{\rm hor}), \\
I_{\rm ver}(B,t) = n_{\rm ver}\Gamma_{\rm ver}(B)\exp(-t/\tau_{\rm ver}).\label{eq:Ihorver1_SI}
\end{aligned}
\end{equation}
The factor 2 for horizontal nanoplatelets shows that two orthogonal dipoles contribute to the emission. In contrast, in vertical nanoplatelets only one of these dipoles contributes to the emission. For the time-integrated intensities $I_{\rm hor}^{\rm int}(B)$ and $I_{\rm ver}^{\rm int}(B)$ one finds 
\begin{equation}
\begin{aligned}
I_{\rm hor}^{\rm int}(B) = 2n_{\rm hor}\Gamma_{\rm hor}(B) \tau_{\rm hor}, \\
I_{\rm ver}^{\rm int}(B) = n_{\rm ver}\Gamma_{\rm ver}(B) \tau_{\rm ver}.
\label{eq:Ihorverint_SI}
\end{aligned} 
\end{equation}

The radiative recombination rate of the dark excitons in vertically oriented NPLs increases in magnetic field as 
\begin{eqnarray}
\Gamma_{\rm ver}(B)=\Gamma_{\rm F}^0+\Gamma_{\rm F}^{\rm B}(B,\pi/2) \, , \\
\Gamma_{\rm F}^{\rm B}(B,\theta)= \frac{\sqrt{1+\zeta^2+2\zeta \cos \theta}-1-\zeta \cos \theta}{\sqrt{1+\zeta^2+2\zeta \cos \theta}} \cdot \frac{\Gamma_{\rm A}}{2} \, .
\label{eq:Gver}
\end{eqnarray}
where $\zeta=\mu_{\rm B}g_eB/\Delta E_{\rm AF}$. This expression is valid in both limits of week and strong magnetic fields. In the case $\zeta<<1$ valid up to the 15 T for the parameters of our NPLs, $\Gamma_{\rm F}^{\rm B}(B,\theta)$ can be well described by the simplified experssion \eqref{EQ:gammaFBM}.  It results in a shortening of both $\tau_{\rm M}$ and $\tau_{\rm L}$ with increasing magnetic field.
For the horizontally oriented NPLs one would expect $\Gamma_{\rm hor}=\Gamma_{0}+\Gamma_{\rm db}(1-\rho_{\rm ex}\rho_{\rm db})$, as the magnetic field does not mix bright and dark exciton states in this case. The experimentally observed magnetic field dependence of the PL decay (Fig.~\ref{fig:Fig1ab}b) is well described as result of a shortening of the dark exciton lifetime in the ensemble of vertically aligned NPLs. Due to the anisotropy of the dark exciton $g$ factor, these nanoplatelets should posses zero Zeeman splitting and consequently zero DCP. However, the experimentally observed DCP in Sample~1 reaches 50\% already at $B=3$~T. This fact indicates the presence of NPLs oriented horizontally or slightly tilted with respect to the substrate (see Fig.~\ref{fig:Fig4_theory_main}b).

This contradiction can be resolved, if one assume a strong coupling or energy transfer between the horizontally and vertically oriented NPLs resulting in the same life time for both orientations $\tau_{\rm hor}=\tau_{\rm ver}=\tau$. We introduce the additional radiative recombination rate $\gamma\Gamma_{\rm F}^{\rm B}(B,\pi/2)$ so that $\Gamma_{\rm hor}(B)=\Gamma_{0}+\gamma\Gamma_{\rm F}^{\rm B}(B,\pi/2)+\Gamma_{\rm db}(1-\rho_{\rm ex}\rho_{\rm db})$ and the nonradiative recombination rate $(1-\gamma)\Gamma_{\rm F}^{\rm B}(B,\pi/2)$ in horizontal NPLs where $\gamma$ is a fit parameter. With this assumption the ratio of intensity from horizontal nanoplatelets to the total intensity from the bi-modal ensemble is:
\begin{gather}
A(B)=\frac{I_{\rm hor}}{I_{\rm hor}^{\rm} +I_{\rm ver} }=\frac{2\Gamma_{\rm hor}(B)}{2\Gamma_{\rm hor}(B)+\eta\Gamma_{\rm ver}(B)} \, , 
\label{eq:Ihorverigen_SI}
\end{gather}
where $\eta=n_{\rm ver}/n_{\rm hor}$.
Note that equation~\eqref{eq:Ihorverigen_SI} is valid both for the momentary and the time-integrated intensities. 

\begin{figure}[h!]
 	\includegraphics{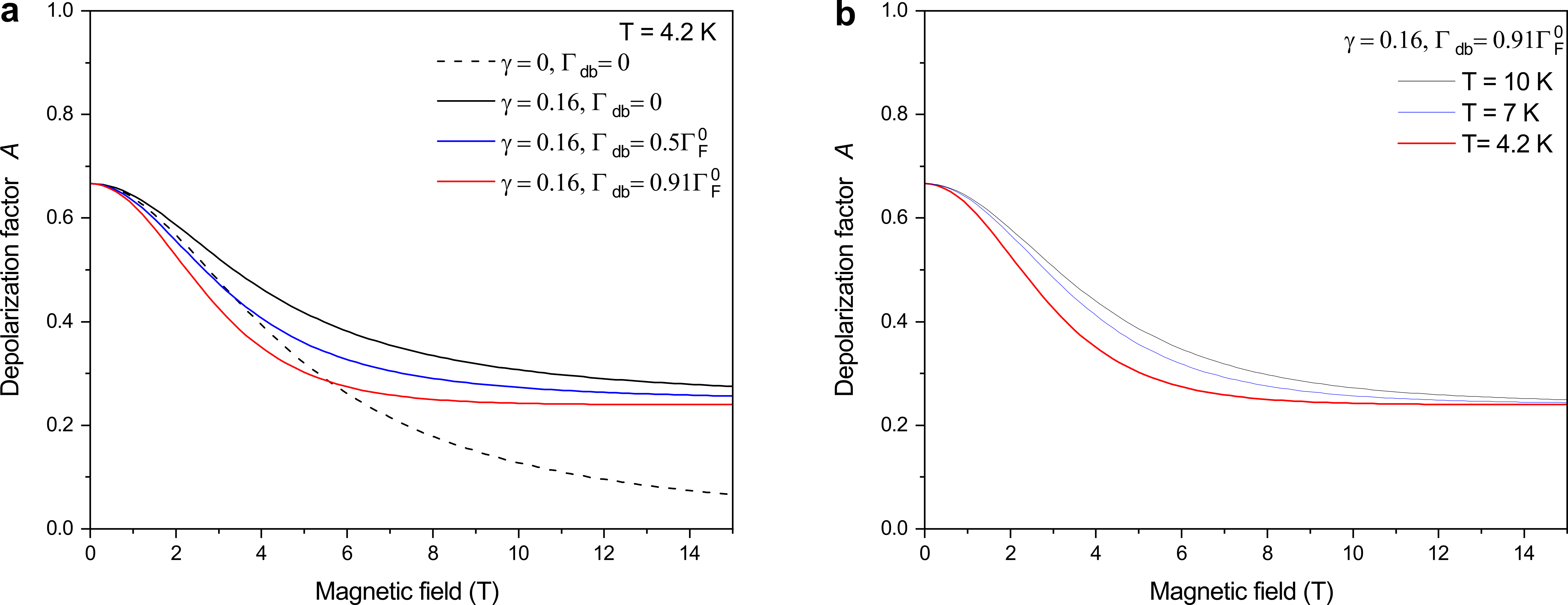}
 	\caption{Dependence of depolarization factor $A(B)$ on  ratio of $\Gamma_{\rm db}/\Gamma_{\rm F}^0$ at  $T=4.2$~K (panel \textbf{a}) and on  temperature (panel \textbf{b}).
 	}
 	\label{fig:def_SI}
 \end{figure}

The dependence of the depolarization factor $A(B)$ on the ratio $\Gamma_{\rm db}/\Gamma_{\rm F}^0$ and the temperature with $\eta=1$ and $\gamma=0.16$ used for fitting the DCP of Sample~1 is shown in figure ~\ref{fig:def_SI}. One can see that the dangling bond assisted recombination results in a decrease of $A(B)$. The thermal depolarization of the dangling bonds suppresses the dangling bond assisted recombination so that $A(B)$ increases (see Fig.~\ref{fig:def_SI}b) .
 
\subsection{Fitting of the equilibrium degree of circular polarization} \label{subsec:SI_Theor_Fitting_Peq}

\begin{figure}[h!]
	\includegraphics{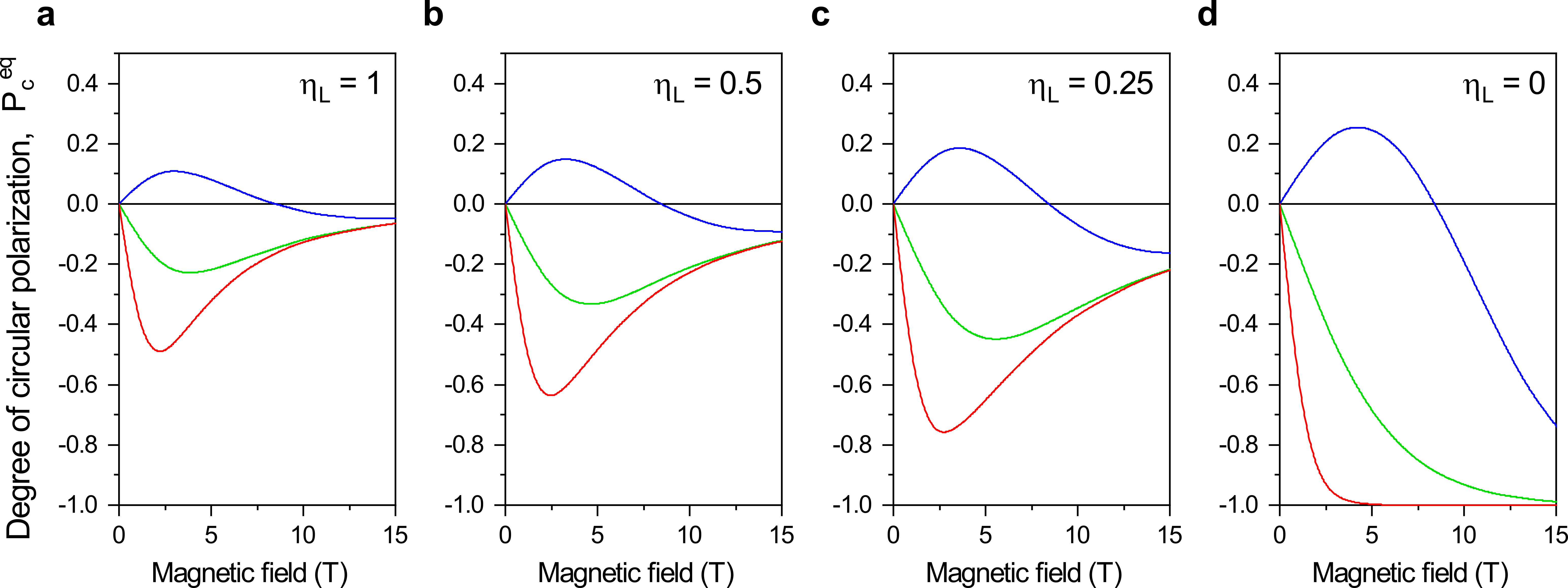}
	\caption{Degree of circular polarization $P_{\rm c}^{\rm eq}(B)$ calculated for ensemble of NPLs with $\gamma=0$, ${ g}_{\rm F}=2$, $\Gamma_{\rm db}=0$ and different values of $E_{\rm p}=1.2$(red), $0$ (green) and $-0.55$~meV (blue) and various $\eta_{\rm L}=\eta_{\rm ver}/\eta_{\rm hor}$. $T=4.2$~K.
	}
	\label{fig:eta_SI}
\end{figure}

As it wasshown from fitting of the PL decay (see Fig.~\ref{fig:Fig2abcd}c), the equilibrium degree of circular polarization $P_{\rm c}^{\rm eq}(B)$ is determined by the recombination and spin polarization of the dark excitons in the NPLs with lifetime $\tau_{\rm L}$. Equation \eqref{eq:PcB_SI} can be rewritten for horizontal and vertical NPLs with lifetime $\tau_{\rm L}$ as:
\begin{gather}
P_{\rm c}^{\rm eq}(B)=\frac{I_{\rm hor}^{\rm L}}{I_{\rm hor}^{\rm L}+I_{\rm ver}^{\rm L}}P_{\rm c,L}^{\rm hor}(B)=\frac{2\Gamma_{\rm hor}(B)}{2\Gamma_{\rm hor}(B)+\eta_{\rm L}\Gamma_{\rm ver}(B)}P_{\rm c,L}^{\rm hor}(B), \\
P_{\rm c,L}^{\rm hor}(B)=\frac{-\rho_{\rm ex}(\Gamma_{\rm F}^0+\gamma\Gamma_{\rm F}^{\rm B}(B)+\Gamma_{\rm db})+\rho_{\rm db}\Gamma_{\rm db}}{\Gamma_{\rm F}^0+\gamma\Gamma_{\rm F}^{\rm B}(B)+\Gamma_{\rm db}(1-\rho_{\rm ex}\rho_{\rm db})},
\end{gather}
The fitting of the experimental dependence of $P_{c}^{\rm eq}(B)$ in Fig.~\ref{fig:Fig2abcd}d is achieved with $g_{\rm F}=g_e-3g_h=2$, $E_{\rm p}=1.2$~meV, $\Gamma_{\rm db}=0$, $\eta_{\rm L}=1$, and $\gamma=0.16$. The introduction of the exchange energy $E_{\rm p}=1.2$~meV is necessary to describe the temperature dependence of the DCP increase in weak magnetic fields. The parameters $\eta_{\rm L}$ and $\gamma$ are responsible for the decrease of DCP in high magnetic fields. The parameter $\eta_{\rm L}$ also influences the maximum achievable DCP in magnetic fields $B<5$~T as shown in Fig.~\ref{fig:eta_SI}. The choice $\eta_{\rm L}=1$ gives the best fit of the experimental data for $P_{\rm c}^{\rm eq}(B)$ for all temperatures.

\begin{figure}[h]
	\includegraphics[width=0.8\linewidth]{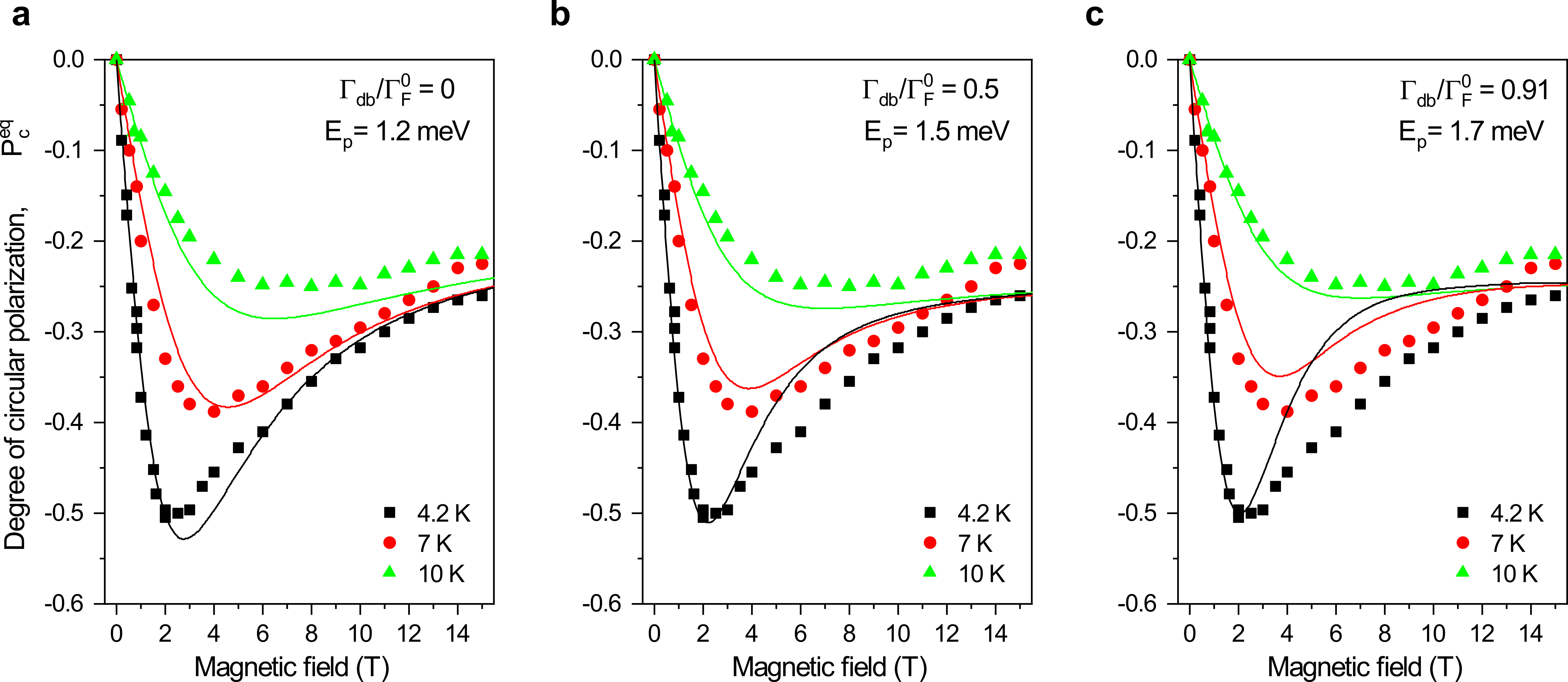}
	\caption{Modeling of equilibrium DCP for CdSe NPLs (Sample 1). \textbf{a},  $\Gamma_{\rm db}/\Gamma_{\rm F}^0=0$ (identical to Fig.~\ref{fig:Fig2abcd}d), \textbf{b}, $\Gamma_{\rm db}/\Gamma_{\rm F}^0=0.5$, and \textbf{c}, $\Gamma_{\rm db}/\Gamma_{\rm F}^0=0.91$. Experimental data are shown by symbols and are identical to those in Fig.~\ref{fig:Fig2abcd}d. 
	}
	\label{fig:SIsample1Gdb}
\end{figure}

Fitting of the DCP for Sample~1 in Fig.~\ref{fig:Fig2abcd}d was done without spin-dependent recombination of the dark exciton. The modeling of the experimental data for the case $\Gamma_{\rm db}/\Gamma_{\rm F}^0=0.5$ and $\Gamma_{\rm db} /\Gamma_{\rm F}^0=0.91$ is shown in Fig.~\ref{fig:SIsample1Gdb}. One can see that the assumption of $\Gamma_{\rm db} \ne 0$ in the case of Sample 1 has no influence on the DCP behavior in low magnetic fields. In high magnetic fields the increase of the $\Gamma_{\rm db} /\Gamma_{\rm F}^0$ ratio results in crossing of the DCP dependencies calculated for different temperatures. These results show that the rate of spin-dependent recombination in the case of Sample 1 is negligibly small compared to the phonon-assisted recombination rate of the dark exciton.   

\newpage
\subsection{Fit of the time-integrated degree of circular polarization} \label{subsec:SI_Theor_Fitting_Pint}

In this part we explain the fitting of the magnetic field dependence of the time-integrated DCP $P_{\rm c}^{\rm int}(B)$ shown in the upper panel of Fig.~\ref{fig:Fig2abcd}d.  In  Sample 1, $|P_{\rm c}^{\rm int}|<|P_{\rm c}^{\rm eq}|$  (compare results inthe  upper and lower panels of Fig.~\ref{fig:Fig2abcd}d). This relation is caused by the different equilibrium DCP in the M and L subensembles, as demonstrated in the Supplementary Section S3B (see Fig.~\ref{fig:SI_TR(B)}d).  

Considering the contributions to the PL signal from the M and L subensembles, we can write the following equation for the time-integrated DCP:
\begin{gather}
P_{\rm c}^{\rm int}(B)=\frac{(I^+-I^-)_{\rm L}+(I^+-I^-)_{\rm M}}{(I^++I^-)_{\rm L}+(I^++I^-)_{\rm M}}=\frac{P_{\rm c,L}(B) I_{\rm L}+P_{\rm c,M}(B)I_{\rm M}}{I_{\rm L}+I_{\rm M}}=\frac{P_{\rm c,L}(B) +P_{\rm c,M} (B)I_{\rm M}/I_{\rm L}}{1+I_{\rm M}/I_{\rm L}} ,
\label{eq:pcintSI}
\end{gather}
where $I_{\rm M}$ and $I_{\rm L}$ are the time-integrated PL intensities from the two subensembles with characteristic lifetimes $\tau_{\rm M}$ and $\tau_{\rm L}$, respectively. The contribution from the fast decay component to $P_{\rm c}^{\rm int}$ can be neglected because of its negligibly small time-integrated intensity $I_{\rm short}=A_{\rm short}\tau_{\rm short}$.

Using the ratio of time-integrated intensities $I_{\rm M}/I_{\rm L}\approx[0.1+(B/20)^2]/[0.9-(B/20)^2]$  (see Supplementary Section~\ref{subsec:SI_TR(B)_Theor}), as obtained from a three-exponential fit for Sample 1 at $T=4.2$~K, we can rewrite the equation \eqref{eq:pcintSI} as
\begin{gather}
P_{\rm c}^{\rm int}(B)=P_{\rm c, M}(B)\left[0.1+(B/20)^2\right]+P_{\rm c, L} (B)\left[0.9-(B/20)^2\right].
\end{gather}
The time-integrated polarizations $P_{\rm c, M}(B)$ and $P_{\rm c, L}(B)$ are defined as follows: 
\begin{gather}
P_{\rm c,M}(B)=\frac{2\Gamma_{\rm hor}}{2\Gamma_{\rm hor}+\eta_{\rm M}\Gamma_{\rm ver}}P_{\rm c,M}^{\rm hor}(B), \\
P_{\rm c,M}^{\rm hor}(B)=\frac{-\rho_{\rm ex, M}(\Gamma_{\rm F}^0+\gamma_{\rm M}\Gamma_{\rm F}^{\rm B}(B,\pi/2)+\Gamma_{\rm db})+\rho_{\rm db}\Gamma_{\rm db}}{\Gamma_{\rm F}^0+\gamma_{\rm M}\Gamma_{\rm F}^{\rm B}(B,\pi/2)+\Gamma_{\rm db}(1-\rho_{\rm ex}\rho_{\rm db})}, \label{eq:Pcmhor}\\
P_{\rm c,L}(B)=\frac{2\Gamma_{\rm hor}}{2\Gamma_{\rm hor}+\eta_{\rm L}\Gamma_{\rm ver}}P_{\rm c,L}^{\rm hor}(B), \\
P_{\rm c,L}^{\rm hor}(B)=\frac{-\rho_{\rm ex, L}(\Gamma_{\rm F}^0+\gamma_{\rm L}\Gamma_{\rm F}^{\rm B}(B,\pi/2)+\Gamma_{\rm db})+\rho_{\rm db}\Gamma_{\rm db}}{\Gamma_{\rm F}^0+\gamma_{\rm L}\Gamma_{\rm F}^{\rm B}(B,\pi/2)+\Gamma_{\rm db}(1-\rho_{\rm ex}\rho_{\rm db})} . \label{eq:Pclhor}
\end{gather}
Note that in equations (\ref{eq:Pcmhor}, \ref{eq:Pclhor}) we use the equilibrium exciton polarizations $\rho_{\rm ex, M}$ and $\rho_{\rm ex, L}$, as the spin relaxation time is much shorter than the decay times $\tau_{\rm M}$ and $\tau_{\rm L}$. 
According to equation~\eqref{eq:PcB_SI}, a smaller DCP in the subensemble with $\tau_{\rm M}$ suggests two possibilities. First, in this subensemble the Zeeman splitting in the horizontal NPLs is smaller compared to the L subensemble, e.g., due to a different exchange interaction energy with the DBSs. This might be caused by presence of surface traps in the M subensemble, resulting in nonradiative recombination and in suppression of dangling bond formation. The
Fitting of the experimental dependencies for $P_{\rm c}^{\rm int}(B )$ as presented in Fig.~\ref{fig:Fig2abcd}c was performed with $\eta_{\rm M}=1.3$,$\eta_{\rm L}=1$, $g_{\rm F}=2$, $\Gamma_{\rm db}=0$, $\gamma_{\rm M}=\gamma_{\rm L}=0.16$, $E_{\rm p}=1.2$~meV for the subensemble with $\tau_{\rm L}$ and $E_{\rm p}=0$ meV for the subensemble with $\tau_{\rm M}$.  The rates $\Gamma_{\rm F}^0$ and $\Gamma_{\rm F}^{\rm B}(B,\pi/2)$ determined from fitting of the time-resolved PL in Sample 1 were used thereby (see Fig.~\ref{fig:SI_TR(B)}a).

\begin{figure}[t]
	\includegraphics{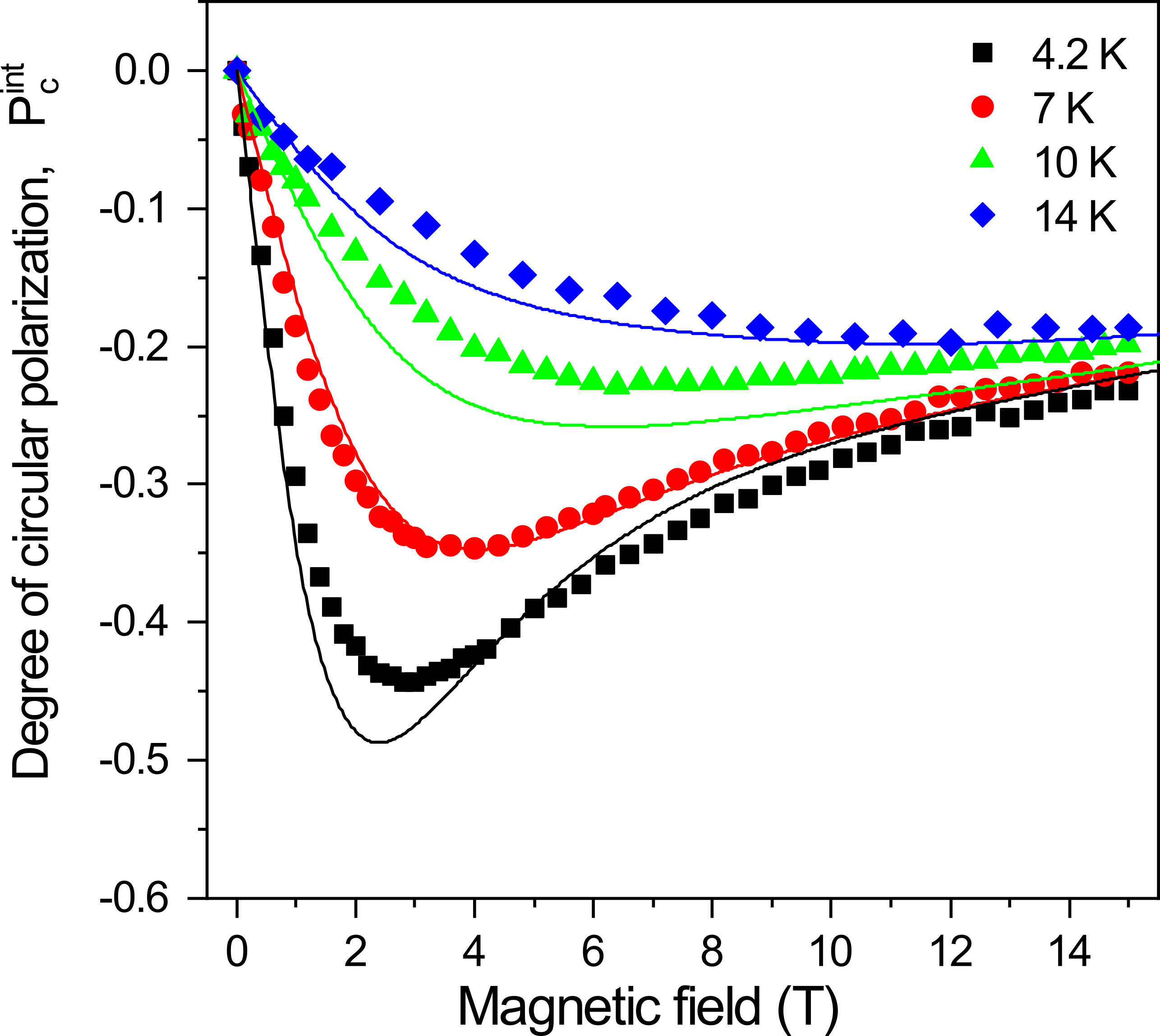}
	\caption{Magnetic field dependence of time-integrated DCP in 4ML CdSe NPLs (Sample 1) at various temperatures. Symbols are experimental data from upper part of Fig.~\ref{fig:Fig2abcd}d. Lines show model calculations of $P_{\rm c}^{\rm int}(B)$ for ensemble of NPLs with $\eta_{\rm M}=3.5$, $\eta_{\rm L}=1$, $\gamma_{\rm L}=0.16$, $\gamma_{\rm M}=0.4$, $g_{\rm F}=2$, $\Gamma_{\rm db}=0$ and $E_{\rm p}=1.2$~meV in both subensembles.}
	\label{fig:SIPint}
\end{figure}

Another way to describe the difference between $P_{\rm c}^{\rm eq}$ and $P_{\rm c}^{\rm int}$ is based on a larger number of vertical NPLs and a larger value of the $\gamma$ parameter in the subensemble with lifetime $\tau_{\rm M}$. The result of fitting with this assumption is presented in Fig.~\ref{fig:SIPint} with $\eta_{\rm M}=3.5$, $\eta_{\rm L}=1$, $g_{\rm F}=2$, $\Gamma_{\rm db}=0$, $\gamma_{\rm M}=0.4$, $\gamma_{\rm L}=0.16$, and $E_{\rm p, M}=E_{\rm p, L}=1.2$~meV. Note that in this case the exchange interaction energy of the dark exciton with DBSs is the same for the M and L subensembles. One can see that this approach also gives a good description of the experimental dependencies of the time-integrated DCP in Sample 1. Therefore, the available experimental data do not allow us to identify unequivocally the reason for the difference between the saturated DCP in the subensembles with lifetimes $\tau_{\rm M}$ and $\tau_{\rm L}$.    

\begin{table*}
	\small
	\caption{Notations used in the paper}
	\begin{tabular*}{1\textwidth}{rl}
		Notation & Definition\\ \hline
		$\Gamma_{\rm A}$ - &  recombination rate of bright exciton\\
		$\Gamma_{\rm F}$ -& recombination rate of dark exciton via spin-independent mechanisms  \\
		$\Gamma_{\rm F}^0$ -& radiative recombination rate of the dark exciton in zero magnetic field, $\Gamma_{\rm F}^0= \Gamma_0+\Gamma_{\rm db}$ \\
		$\Gamma_{0}$ -& spin-independent radiative recombination rate of the dark exciton in zero magnetic field \\
		$\Gamma_{\rm F}^{\rm B}$ -& radiative recombination rate of the dark exciton induced by magnetic field\\
		$\Gamma_{\rm db}$ -& radiative recombination rate of the dark exciton in zero magnetic field assisted by interaction with dangling bond spins  \\
				$\Gamma_{\rm hor}$ -& full radiative recombination rate of the dark exciton in horizontal NPLs \\
		$\Gamma_{\rm ver}$ -& full radiative recombination rate of the dark exciton in vertical NPLs \\
		$\Gamma_{\rm hor}^{\pm 2}$ -& radiative recombination rate of $\pm 2$ dark excitons  in horizontal NPLs\\
		$\Gamma_{\pm 2}$ -& full recombination rate of $\pm 2$ dark excitons  in horizontal NPLs \\
		$\Gamma_{\rm nr}$ -& nonradiative recombination rate of the dark exciton, $\Gamma_{\rm nr}=\Gamma_{\rm M}-\Gamma_{\rm L}$\\
		$\Gamma_{\rm L}$ -& long PL decay rate \\ 
		$\Gamma_{\rm M}$ -& middle PL decay rate \\
		$\tau_{\rm L}$ -& long PL decay time	\\ 
     	$\tau_{\rm M}$ -& middle PL decay time	\\
     	$\tau_{\rm short}$ -& short PL decay time (i.e. initial fast decay)	\\
     	$\tau_{\rm s}$ -& spin relaxation time of exciton \\
     	$\Delta E_{\rm AF}$ -& bright-dark exciton splitting \\
     		$\Delta E_{\rm F}$ -& Zeeman splitting of dark exciton \\
     	$E_{\rm p}$ -& exchange interaction energy of exciton with dangling bonds spins \\
     	$\rho_{\rm ex}$ -& spin polarization of dark exciton \\
          	$\rho_{\rm db}$ -&spin polarization of dangling bonds \\
     	$n_{\rm db}$ -&surface density of dangling bonds with unpaired spins \\
        $P_{\rm c}(B)$ -&degree of circular polarization  of PL induced by magnetic field\\
     	$P_{\rm c}(t)$ -& time-dependent degree of circular polarization  of PL \\
     	$P_{\rm c}^{\rm eq}(B)$ -& equilibrium degree of circular polarization  of PL \\
     	$P_{\rm c}^{\rm hor}(B)$ -& degree of circular polarization of PL in horizontal nanoplatelets\\
     	$P_{\rm c}^{\rm int}(B)$ -& time-integrated degree of circular polarization  of PL \\		
     	$P_{\rm c}^{\rm sat}$ -& maximum achievable degree of circular polarization  of PL (saturation level)\\
     	$g_{\rm e}$ -& electron $g$-factor\\
     	$g_{\rm h}$ -& hole $g$-factor\\	
     	$g_{\rm F}$ -& $g$-factor of the dark exciton\\	
     	$g_{\rm {db}}$ -& $g$-factor of the DBS\\	
        $\alpha_{j}$ -& exchange constant for interaction with $j$-th dangling bond spin with electron spin\\
     	$\alpha$ -& average exchange constant \\
     	$\eta$ -& ratio of vertical to horizontal NPLs\\		   		
 		\hline
	\end{tabular*}
	\label{tab:table2}
\end{table*}


\begin{thebibliography}{99}
\bibitem{Kalitukha2018} The electron $g$-factor for 4ML CdSe NPLs was measured by spin-flip Raman scattering at $T=2$~K. It is anisotropic being  $g^{\parallel}_{\rm e}=1.60$ for the direction normal to the NPL plane (i.e. along the $z$ axis) and $g^{\perp}_{\rm e}=1.78$ for the in-plain directions. The results of these measurements performed by  V. F. Sapega and I. V. Kalitukha will be published elsewhere. For model evaluations we took the averaged value of  $g_{\rm e}=1.70$.
\end{thebibliography}
\end{document}